\documentclass{aa}  

\usepackage{xcolor}
\usepackage{soul}
\sethlcolor{yellow}

\def\Lir{$L_{\textrm{IR}}/L_{\ast}$}
\def\gaia{\textit{Gaia}}
\def\hipp{\textit{Hipparcos}}

\newcommand{\diskincl}{${75.7}^{+1.1}_{-1.3}$$\deg$}
\newcommand{\diskrad}{${118\pm9}\textrm{au}$}

\renewcommand{\deg}{^\circ}

\def\apjl{ApJL}

\def\apjs{ApJS}

\def\apjs{ApJS}

\def\aap{A\&A}

\usepackage{graphicx}
\usepackage{txfonts}

\usepackage{hyperref}
\hypersetup{nolinks=true} 

\begin{document} 

   \title{The first scattered light images of HD\,112810, a faint debris disk in the Sco-Cen association}

   \author{E. C. Matthews\inst{\ref{mpia},\ref{obsge}}
          \and
          M. Bonnefoy\inst{\ref{grenoble}}
          \and
          C. Xie\inst{\ref{lam}}
          \and
          C. Desgrange\inst{\ref{mpia},\ref{grenoble}}
          \and
          S. Desidera\inst{\ref{inaf}}
          \and
          P. Delorme\inst{\ref{grenoble}}
          \and
          J. Milli\inst{\ref{grenoble}} 
          \and
          J. Olofsson\inst{\ref{mpia}}
          \and
          D. Barbato\inst{\ref{obsge}}
          \and
          W.~Ceva\inst{\ref{obsge}}
          \and
          J.-C.~Augereau\inst{\ref{grenoble}}
          \and
          B. A. Biller\inst{\ref{edinburgh}}
          \and
          C. H. Chen\inst{\ref{stsci}}
          \and
          V. Faramaz-Gorka\inst{\ref{steward}} 
          \and
          R. Galicher\inst{\ref{lesia}}
          \and
          S. Hinkley\inst{\ref{exeter}}
          \and
          A.-M. Lagrange\inst{\ref{lesia}}
          \and
          F. M\'enard\inst{\ref{grenoble}}
          \and
          C. Pinte\inst{\ref{monash},\ref{grenoble}}
          \and
          K. R. Stapelfeldt\inst{\ref{jpl}}
          }

   \institute{
        Max-Planck-Institut f\"ur Astronomie, K\"onigstuhl 17, D-69117 Heidelberg, Germany \\ \email{matthews@mpia.de}\label{mpia}
        \and
        Department of Astronomy, University of Geneva, Chemin Pegasi 51, CH-1290 Versoix, Switzerland \label{obsge}
        \and
        Universit\'e Grenoble Alpes, CNRS, IPAG, F-38000 Grenoble, France\label{grenoble}
        \and
        Aix Marseille Univ, CNRS, CNES, LAM, Marseille, France \label{lam}
        \and
        INAF-Osservatorio Astronomico di Padova, Vicolo dell'Osservatorio 5, I-35122 Padova, Italy\label{inaf}
        \and
        Scottish Universities Physics Alliance, Institute for Astronomy, University of Edinburgh, Blackford Hill, Edinburgh EH9 3HJ, UK; Centre for Exoplanet Science, University of Edinburgh, Edinburgh EH9 3HJ, UK\label{edinburgh}
        \and
        Space Telescope Science Institute, 3700 San Martin Drive, Baltimore, MD 21218, USA\label{stsci}
        \and
        Steward Observatory, Department of Astronomy, University of Arizona, 933 N. Cherry Ave, Tucson, AZ 85721, USA\label{steward} 
        \and
        LESIA, Observatoire de Paris, Université PSL, CNRS, Université Paris Cité, Sorbonne Université, 5 place Jules Janssen, 92195 Meudon, France\label{lesia}
        \and
        University of Exeter, Astrophysics Group, Physics Building, Stocker Road, Exeter, EX4 4QL, UK.\label{exeter}
        \and
        School of Physics and Astronomy, Monash University, Vic 3800, Australia\label{monash}
        \and
        Jet Propulsion Laboratory, California Institute of Technology, Mail Stop 321-100, 4800 Oak Grove Drive, Pasadena CA 91109 USA\label{jpl}
        \\
            }

   \date{Received --; accepted --}
   \authorrunning{Matthews et al.}
   \titlerunning{HD\,112810 scattered light disk detection}

  \abstract{Circumstellar debris disks provide insight into the formation and early evolution of planetary systems. Resolved belts in particular help to locate planetesimals in exosystems, and can hint at the presence of disk-sculpting exoplanets.}{We study the circumstellar environment of \object{HD\,112810} (HIP\,63439), a mid-F type star in the Sco-Cen association with a significant infrared excess indicating the presence of a circumstellar debris disk.}{We collected five high-contrast observations of HD\,112810 with VLT/SPHERE. We identified a debris disk in scattered light, and found that the debris signature is robust over a number of epochs and a variety of reduction techniques. We modelled the disk, accounting for self-subtraction and assuming that it is optically thin.}{We find a single-belt debris disk, with a radius of \diskrad~and an inclination angle of \diskincl. This is in good agreement with the constraints from SED modelling and from a partially-resolved ALMA image of the system. No planets are detected, though planets below the detection limit ($\sim$2.6M$_\textrm{J}$~at a projected separation of 118au) could be present and could have contributed to sculpting the ring of debris.}{HD\,112810 adds to the growing inventory of debris disks imaged in scattered light. The disk is faint, but the radius and the inclination of the disk are promising for follow-up studies of the dust properties.}


   \maketitle
%

\section{Introduction}

The outer reaches of exoplanet systems remain difficult to study, with only $\sim$100 exoplanets detected beyond $\sim$10au to date\footnote{See \href{https://exoplanetarchive.ipac.caltech.edu/}{https://exoplanetarchive.ipac.caltech.edu/}}. Young exoplanet systems are particularly difficult to study since transit and radial velocity surveys are generally less sensitive to planets around young, active stars -- only $\sim$5\% of planets with known ages are younger than 100Myr. However, understanding the properties of young planets at all spatial scales, and their population statistics, is key to understanding the formation and early evolution of exoplanetary systems.

Circumstellar debris disks provides a unique view of these outer exosystems \citep[for a review see][]{hughes2018} 
. While protoplanetary disks are the sites of active ongoing planet formation, slightly older star systems are often surrounded by circumstellar debris disks that are made up of planetesimals and small dust grains in orbit around the star. The small particles are second-generation dust that is generated by collisions of the planetesimals \citep[e.g.][and references therein]{wyatt2008}. Detection of these small particles therefore implies that there is a reservoir of planetesimals, and hints at the location of this material. Further, this dust is quickly depleted from the system, so observations of the small dust indicate that these collisions between planetesimals are ongoing: this could imply the presence of unseen planets, that stir the disk and enhance the rate of collisions; such unseen planets would also impact the structure of the disk \citep[e.g][]{wyatt1999,krivov2010,matthews2014}. In many cases, planets would require masses between that of Neptune and Jupiter to explain the level of stirring observed in well characterized debris disks \citep{pearce2022}.

The presence of debris disks can be inferred from a Spectral Energy Distribution (SED) in the case that significant excess emission in the mid- and far-infrared relative to the predicted emission of a stellar photosphere alone. If the excess is well characterized, the disk temperature(s) can also be inferred from the SED alone, and with suitable assumptions about the grain properties, corresponding disk location(s) can be derived. Some nearby, bright disks can also be resolved, allowing for detailed study of their morphology. Several debris disks have been resolved in thermal emission since the first such detections with \textit{IRAS} \citep{aumann1985}, and in scattered light in the optical and near-IR with high contrast imaging facilities \citep{smith1984,schneider1999}. In the last decade, both scattered light and thermal disk imaging has advanced significantly. The advancement of adaptive optics (AO) and coronagraphic technology has led to many new disks being detected in scattered light optical- and near-IR high-contrast imaging \citep[e.g.][]{kasper2015,currie2015,lagrange2016,matthews2017,bonnefoy2021}; ALMA has provided an unprecedented look at the thermal emission of disks at millimeter wavelengths \citep[e.g][]{boley2012,macgregor2013,ricci2015,liemansifry2016,macgregor2019}. These thermal and scattered light images allow for positions of planetesimal belts and small dust grains in the system to be traced, and also open the opportunity for detailed characterization of the grain properties, such as measuring the scattering phase function of the disk and thereby constraining grain sizes and compositions \citep[e.g.][]{milli2017}. 

This treasure trove of resolved thermal and scattered light images reveals complex structure that cannot be deduced from the SED alone, such as spirals, warps and gaps between distinct belts of dust \citep{feldt2017,bonnefoy2017,marino2018,marino2019,macgregor2019}. One proposed explanation of such complex disk features is the gravitational influence of unseen planets in the system. More generally, debris disks may provide an insight into the bulk compositions of exoplanets, since the debris material is generated in collisions of planetesimals -- which are also the building blocks for any rocky planets forming at the widest separations.

The Scorpius-Centaurus OB association has been a particularly fruitful location for studies of young circumstellar disks. Many of its debris disks, as well as some protoplanetary and transition disks, have been studied in detail in this region. The association is young, with members spanning $\sim$7-18Myr \citep{pecautmamajek2016}, is nearby ($\sim$110-130pc), and has a disk fraction of around a third, with no significant variation in disk fraction between the three sub-groups \citep{chen2011}. As a large association with a relatively narrow age range, the region is particularly valuable since it provides a snapshot of the outer architectures in planetary systems, even among stars with similar masses and formation environments. A number of scattered light debris disk studies with VLT/SPHERE and Gemini/GPI have revealed significant diversity of disk architectures among Sco-Cen stars, hinting at the possible presence of many outer planetary systems \citep[e.g.][]{kalas2015,hung2015,feldt2017,bonnefoy2017,matthews2017,engler2020,hom2020,bonnefoy2021}.

\object{HD\,112810} (\object{HIP\,63439}) is an F4 type star in the Lower Centaurus Crux (LCC) region of the Scorpius-Centaurus association \citep{dezeeuw1999,rizzuto2011,gagne2018}, with an age of $\sim$17Myr \citep{pecaut2012} and a distance of 133.7$\pm$0.3pc \citep{gaia_edr3}. The star has a well-characterized infrared excess, which has been studied with several facilities including \textit{Spitzer} \citep{werner2004} and \textit{WISE} \citep{wright2010}. 
\citet{chen2011} detected excess emission at both 24$\mu$m and 70$\mu$m with Spitzer MIPS, and \citet{chen2014} subsequently presented Spitzer IRS spectroscopy of the target. HD\,112810 also shows a clear excess in the WISE W3 (11.6$\mu$m) and W4 (22.1$\mu$m) channels \citep{mcdonald2012,cottensong2016}. 
\citet{chen2014} used blackbodies to model the Spitzer MIPS and IRS measurements of HD\,112810, and found evidence for two distinct temperatures of dust, at 258$^{+5}_{-7}$K (\Lir=1.3$\times10^{-4}$) and 57$^{+6}_{-4}$K (\Lir=5.9$\times10^{-4}$) respectively. This corresponds to radii of $\sim$8au and $\sim$160au, following the conversion in \citet{pawellek2015} which modifies the blackbody equilibrium temperature of grains based on realistic compositions, though there is an inherent uncertainty in deriving dust positions from SED information alone. \citet{jangcondell2015} modelled the same Spitzer data as presented in \citet{chen2014} using more realistic grain properties (as opposed to blackbodies), and derived somewhat hotter temperatures of 405$\pm$35K and 69.5$\pm$0.9K, corresponding to physical locations of 0.951$\pm$0.2au and 24.3$\pm$2.9au using the grain property treatment described in \citet{jangcondell2015}. The \Lir~value is towards the faintest end of the distribution of disks that have been detected in scattered light to date \citep[see their Fig.~15]{esposito2020}. 

The HD\,112810 debris disk has been detected in the continuum with ALMA \citep{liemansifry2016} at a peak signal to noise ratio of 8.6. The disk was resolved along the major but not minor axis, leading the authors to derive a disk inclination $>$67$\deg$, and an outer radius for the dust belt of 130$^{+80}_{-70}$au (using the updated \gaia~distance to the target). Only relatively edge-on disks are detectable in total intensity scattered light images, for several reasons. For an optically thin disk, the integrated dust quantity along each line of sight is higher for a highly inclined disk, and this corresponds to a higher peak disk flux. Many disks are strongly forward scattering \citep[e.g.]{milli2017}, further boosting the peak surface brightness of an edge-on disk relative to a face-on disk. Further, the angular differential imaging technique for high contrast imaging is biased against detecting axisymmetric signal: in this method, a sequence of observations at different telescope orientations are used to model and subtract the stellar PSF from each image. Such signal is present regardless of the telescope orientation, and is therefore subtracted with the stellar PSF \citep[e.g.][]{milli2012}. These biases are clearly confirmed in observations: in the sample from \citet{esposito2020}, 15/18 disks detected in total intensity have an inclination $\geq$80$\deg$, and 17/18 have inclination $\geq$75$\deg$. 

The ALMA observations of HD\,112810, which indicate a relatively large outer dust radius and an inclination of >67$\deg$, suggest that a total intensity scattered light detection of the disk may be possible, but is challenging depending on the true inclination of the disk. While the most edge-on disks are the easiest to detect, those at more modest inclinations provide interesting characterization opportunities: in particular, the geometry of these disks is such that the front and back edges are well separated from each other and from the host star in disk images, meaning that the scattering phase function (SPF) of the disk can be measured across almost the full range of scattering angles. This can be key in untangling the grain structure and composition \citep[e.g.][]{milli2017}. These grains provide a route to understand the bulk compositions of exoplanets, since these dust grains are generated through the collisions of planetesimals. Such planetesimals should trace the compositions of any rocky exoplanets that form at wide separations.
 
In this paper, we present the first scattered light images of the HD\,112810 debris disk, collected with the VLT/SPHERE high contrast imager. We observed the target as part of two surveys of multi-belt, or `holey' debris disks, selected based on the target SED \citep[][and references therein]{matthews2018,bonnefoy2021}. The twin goals of these surveys were to search for planets in the debris disk gaps, and to characterize the disks themselves in scattered light. The paper is organized as follows: we first reassess the stellar properties of HD\,112810 in Sect.~\ref{sec:star}. We describe our observations and data reduction in Sects.~\ref{sec:obs} and \ref{sec:dr} respectively. We model the disk structure in Sect.~\ref{sec:model} and discuss the constraints on possible planets in the system in Sect.~\ref{sec:planetlimits}. Finally, we discuss key takeaways of this work in Sect.~\ref{sec:discussion}.

\section{Stellar Properties of HD\,112810}
\label{sec:star}

The characterization of the stellar properties of HD\,112810 is based on the methods described in \cite{desidera2021}. HD\,112810 is a single star, with no physical companions detected in the HARPS radial velocity monitoring \citep[16 epochs over 671 days, see Sect.~\ref{sec:detlim-rv};][]{grandjean2023},  the Gemini/NICI snapshot survey \citep{janson2013} or in the \gaia~catalog (where no comoving objects are found up to more than 20000au projected separation\footnote{We searched the \gaia~catalog out to a separation of 180'' (=24,000au at the distance of HD112810). No sources with 5-parameter solutions have both parallax consistent with HD112810 to within 5$\sigma$, and proper motion consistent with HD112810 to within 5$\sigma$.}). Several indicators in the \gaia~data further indicate that unseen stellar companions are unlikely (see Sect.~\ref{sec:detlim-hgca} for details). Therefore, we can safely adopt the \gaia~parameters (DR3) to evaluate the Sco-Cen membership as previously inferred by \citet{dezeeuw1999,rizzuto2011}. Using \mbox{BANYAN $\Sigma$} \footnote{\url{http://www.exoplanetes.umontreal.ca/banyan/banyansigma.php}}\citep{gagne2018}, which compares the \gaia~position, parallax, proper motion and radial velocity of a target to confirmed members of various moving groups, we found a 98.9\% membership probability for LCC, 0.9\% for UCL and just 0.2\% for a field object. Therefore, we assume the membership to LCC and adopt a corresponding age of 17$\pm$3 Myr for the system \citep{pecaut2012}, though we note that future works with \gaia~data will likely determine more precise ages for the Sco-Cen subgroups \citep[see e.g.][]{ratzenboeck2023}.

The star is a mid-F type star with marginally significant reddening. From the observed photometric colours, and adopting E(B-V)=0.016$\pm$0.016 (\citealt{pic}, based on the reddening maps from \citealt{lallement2018}), we derived $T_\textrm{eff}=6637\pm 120$ K using \citet{pecautmamajek2013} tables \footnote{Updated values at \url{https://www.pas.rochester.edu/~emamajek/EEM_dwarf_UBVIJHK_colors_Teff.txt}}. Adopting these values and the \gaia~trigonometric parallax, and limiting the fit to the plausible ages for LCC, we used the PARAM web interface\footnote{ \url{http://stev.oapd.inaf.it/cgi-bin/param_1.3}} \citep{param} to derive the stellar mass and radius. PARAM allows for probabilistic fitting of $T_\textrm{eff}$, [Fe/H], parallax and age to the PARSEC stellar models \citep{bressan2012}, in order to derive stellar mass and radius. For the case of HD\,112810, we used a tight age constraint (17$\pm$3 Myr) corresponding to the Sco-Cen membership, and determined a mass of 1.399$\pm$0.037~$M_{\odot}$ and a radius of 1.369$\pm$0.032~$R_{\odot}$ for HD\,112810. Note that these uncertainties are model-fitting uncertainties only, and do not include any systematic uncertainties in the PARSEC models. The corresponding stellar luminosity is 3.30~$L_{\odot}$.

\section{Observations}
\label{sec:obs}

We imaged HD\,112810 with the SPHERE high contrast imager \citep{beuzit2019} at the VLT. We collected a total of 6 distinct observing sequences on the target, with four of those being successfully completed; details of each observation are given in Table \ref{tab:obslog}. All of our observations were collected in the IRDIFS mode, whereby light is split through a dichroic mirror, and images are simultaneously recorded in two different wavelength ranges, with the dual-band infrared imager IRDIS \citep{dohlen2008,vigan2010,zurlo2014} and the integral field spectrograph \citep[IFS;][]{claudi2008,mesa2015}. To suppress starlight we used the \texttt{N\_ALC\_YJH\_S} coronagraph, which has an inner working angle of $\sim$0.1\arcsec. We collected all of the observations in pupil-stabilized mode, whereby the field naturally rotates with respect to the pupil position over the course of the observing sequence. This means that the stellar speckles have a consistent orientation between images, while any circumstellar material appears to move between images as the field rotates. This allows for angular differential imaging \citep[ADI;][]{marois2006} to construct reference images of the quasi-static stellar speckles that can then be removed from the science images.

The first series of observations, in 2016, were collected as part of three SPHERE programs, all of which had a primary aim of searching for giant planets in systems with known debris disks \citep[see][]{matthews2018,lombart2020}. These observations all used the dual-band DB\_H23 IRDIS filter ($\lambda$=1593nm, $\Delta\lambda$=52nm and $\lambda$=1667nm, $\Delta\lambda$=54nm), and the YJ mode of the IFS (39 wavelength channels between 0.95-1.35$\mu$m), to be optimized for sensitivity to exoplanets. Our data processing revealed a very faint scattered light disk in the IRDIS data (see Sect.~\ref{sec:dr}).

After the identification of this disk, we revisited the target with the aim of boosting the signal-to-noise of our detection. We collected two observation sequences on consecutive nights in February 2020. For these observations, we used the broadband B\_H filter ($\lambda$=1625nm, $\Delta\lambda$=290nm) in order to collect more disk photons, as well as the same YJ mode of the IFS. We also used an extended integration time and field rotation (see Table \ref{tab:obslog}) to improve the sensitivity relative to the early observations.

For observation (1) only acquistion frames were collected before the observation was aborted; for observation (5) the sequence was aborted after 10th cube (140th image) due to a technical fault. In the final cube before the sequence was aborted, the header value \texttt{HIERARCH ESO TEL PARANG END} is incorrectly listed as 0.000, but accurate parallactic angles can still be derived for each slice based on the telemetric values in the header. Since the weather conditions were good until this technical fault, we include this full sequence of observations in our analysis. The other four observations were all collected in good weather conditions, and the target was always at airmass $<$1.2; we list the seeing and field rotation of each observation in Table \ref{tab:obslog}.

Each observing sequence included several calibration frames. Specifically, we collected at least one flux calibration frame and one center calibration frame with both IRDIS and IFS for each observation. For the flux frames, the star center was offset from the coronagraph so that the PSF can be accurately measured. For the center frames (also known as waffle frames), two periodic modulations were applied to the deformable mirror such that four satellite spot images were created in a square pattern, equally displaced from the stellar position. This allows the star center position to be reconstructed to an accuracy of $\sim$0.1pix \citep[1.2mas;][]{vigan2016}  behind the coronagraph. The flux and center calibration frames were collected immediately before and/or after the science sequence. For some observations, we also collected sky background frames at the end of observation sequence.

\begin{table*}[t]
    \centering
    \caption{\label{tab:obslog}Observing log for HD\,112810.}
    \footnotesize
    \begin{tabular}{l|lll|ll|ll|ll}
\hline
ID & Date		&  Program ID   & Mode/Filters     & IRDIS & Total t$_\textrm{exp}$ & IFS & Total t$_\textrm{exp}$ & Seeing & Field Rotation  \\
\hline
1  & 2016-03-02 & 096.C-0713(C) & DB\_H23+YJ & acq. only                      & - & - & - & - & - \\
2  & 2016-03-16 & 096.C-0713(C) & DB\_H23+YJ  & 32x1x64s & 2048s & 2x16x64s & 2048s  & 1.26-1.95 & 19.1$\deg$\\
3  & 2016-05-01 & 097.C-0949(A) & DB\_H23+YJ & 16x5x32s & 2560s & 10x4x64s & 2560s    & 0.51-0.68 & 23.2$\deg$ \\
4  & 2016-05-15 & 097.C-0060(A) & DB\_H23+YJ & 16x8x32s & 4096s & 16x4x64s & 4096s        & 0.51-1.55 & 36.5$\deg$ \\
5  & 2020-02-17 & 0101.C-0016(A) & B\_H+YJ & 10x14x32s & 4480s & 10x7x64s & 4480s    & 0.34-0.48 & 32.1$\deg$ \\
6  & 2020-02-18 & 0101.C-0016(A) & B\_H+YJ & 16x14x32s & 7168s & 16x7x64s & 7168s    & 0.60-1.06 & 58.8$\deg$ \\
\hline
    \end{tabular}
    \tablefoot{All observations were collected in IRDIFS mode, with the IRDIS and IFS subsystems operating simultaneously. The \textit{t} columns indicate the cumulative exposure time across all images for each epoch. For each instrument we list the number of frames, the number of individual images for each frame (NDIT), and the integration time of each individual image (DIT); we also include the cumulative integration time for each observation. }
\end{table*}

\section{Data Processing}
\label{sec:dr}

\begin{figure*}[t]
    \centering
    \includegraphics[width=1.\textwidth]{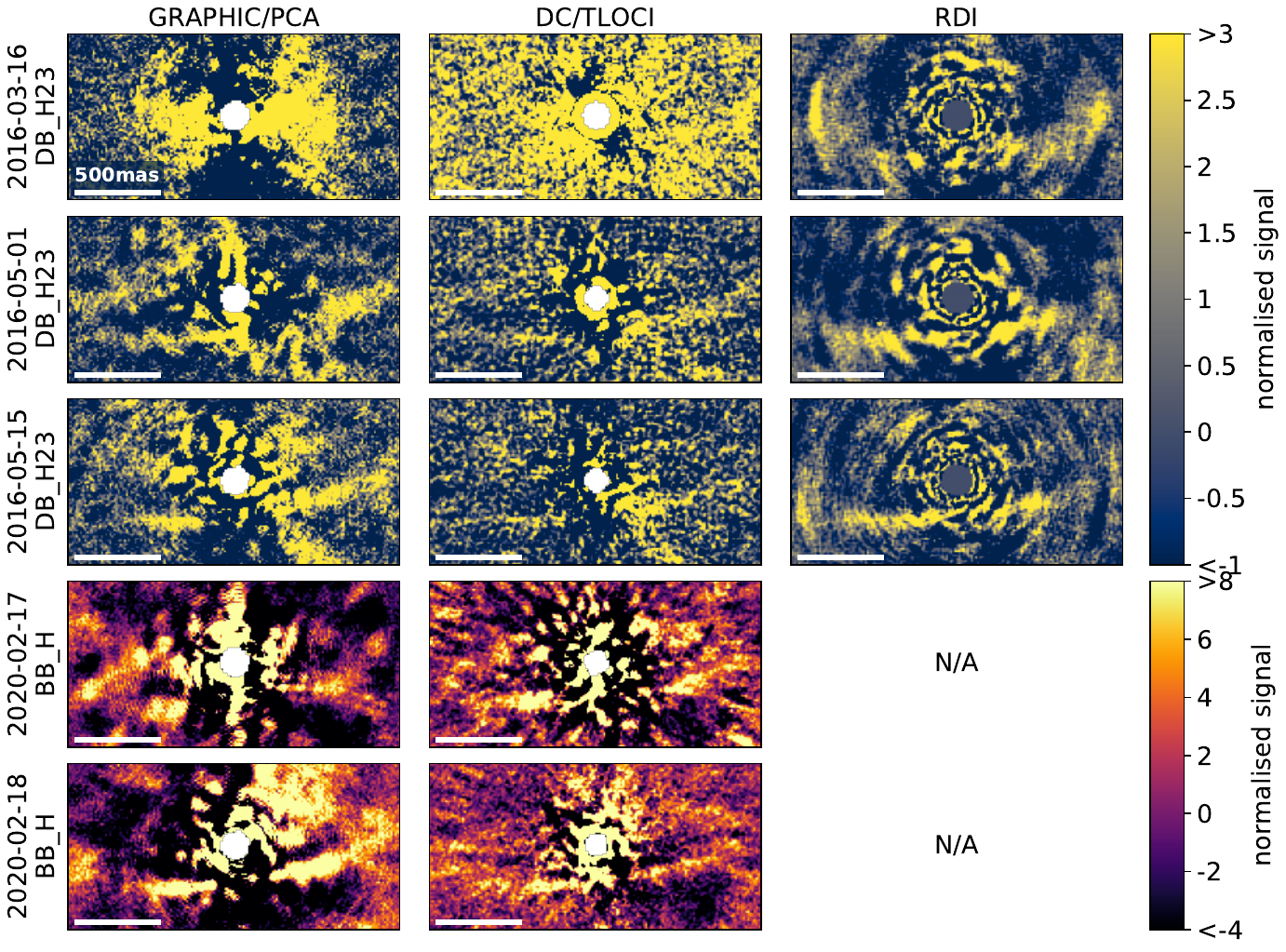}
    \caption{SPHERE/IRDIS images of the HD\,112810 debris disk. Each column indicates data processed via a different method: GRAPHIC/PCA \citep[Sect.~\ref{sec:dr-graphic};][]{hagelberg2016}, TLOCI as provided by the SPHERE Data Center \citep[Sect.~\ref{sec:spheredc};][]{delorme2017} or RDI \citep[Sect.~\ref{sec:rdi};][]{xie2022}; rows indicate each observation epoch. In each case, the two SPHERE/IRDIS channels are co-added. There are no RDI reductions for the BB\_H data, since there is far more data collected with the DB\_H23 filter in the archive, and the reference library we considered does not collate BB\_H data \citep{xie2022}. The `normalised signal' is calculated relative to the background noise in the wide-field; note that this is \textit{not} a true SNR at very small separations, where the noise is higher than the background limit.}
    \label{fig:sphere_irdis_images}
\end{figure*}

The HD\,112810 disk is very faint in scattered light, as can be seen in Fig.~\ref{fig:sphere_irdis_images}. We performed data reductions with several independent pipelines and algorithms, to demonstrate the robustness of our detection and ensure that the extracted physical parameters of the disk are not biased by the specific choice of reduction algorithm. We used two ADI based approaches for each dataset, and also used a reference library RDI based approach for the IRDIS DB\_H23 data. The disk was recovered in all IRDIS data reductions except for the 2016-03-16 epoch, and in none of the IFS data reductions.

The data reduction is split into two distinct phases: firstly a pre-cleaning step, where individual frames are cleaned and calibrated, and secondly a speckle removal step where post-processing algorithms are used to model and remove the stellar PSF and speckle noise. 

\begin{figure*}[t]
    \centering
    \includegraphics[width=\textwidth, trim=0 9.7cm 0 0.cm, clip]{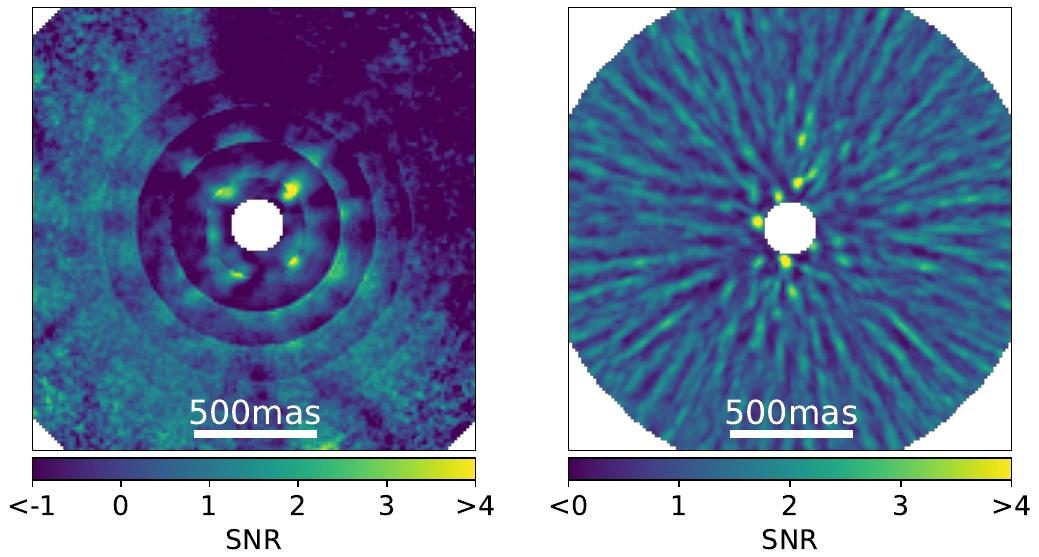}
    \caption{SPHERE/IFS reduced data for HD\,112810. Here we show an image with all data epochs and all wavelengths co-added, with the GRAPHIC/PCA (left) and ANDROMEDA (right) algorithms applied to remove stellar speckles. In both cases, these are SNR maps; for GRAPHIC/PCA each annulus is scaled by the variance in that annulus, causing sharp transitions between concentric annuli. We tried several data reduction approaches and varied the values of tuned parameters. While a few bright spots are seen very close to the inner working angle, these are detected at low significance and are not at consistent locations between epochs -- indicating that these are residual stellar speckles. The disk signal is not visible in any IFS data, either in the individual or in the co-added epochs.}
    \label{fig:sphere_ifs_images}
\end{figure*}

\subsection{ADI with GRAPHIC/PCA}
\label{sec:dr-graphic}

We reduced the SPHERE/IRDIS data from scratch using the GRAPHIC pipeline \citep{hagelberg2016}. GRAPHIC uses a unique Fourier analysis approach to perform frame shifts and rotations, which ensures that high-frequency noise is well preserved during these frame shifts. \citet{hagelberg2016} demonstrate that this technique offers an improved contrast relative to an interpolation method at small separations ($\lesssim$1\arcsec). The pipeline also includes a custom implementation of the commonly-used annular Principal Component Analysis (PCA) approach to model and subtract stellar speckles \citep{amaraquanz2012,soummer2012}.

We briefly describe the key steps here, and refer the reader to \citet{hagelberg2016} for a complete description of the pipeline. Frames are first registered: during this process key information such as the time and parallactic angle of each frame are extracted from the file header, and the position and extent of the stellar PSF are fitted. A master-sky (where available) or master-dark, and a master-flat are applied to the data. Bad pixels are identified from the dark or sky frames using $\sigma$-clipping, and replaced with the median value of neighbouring pixels. Frames are re-centered to align with the waffle frame, and the position of the star behind the coronagraph is extracted from the center calibration frames.

We applied ADI with a PCA algorithm to concentric annuli of the image. This allows for the stellar speckles (which have a consistent orientation between images) to be distinguished from any disk/planet signal (which moves between images as the field rotates). When calculating the PCA components to subtract from each frame, we use a minimum separation criterion in parallactic angle that varies with radial separation so as to minimize self-subtraction. Several PCA components are subtracted from each of the images to remove the stellar speckle and leave disk/planet signal. These subtracted frames are finally derotated to the same on-sky angle, and combined. This process is carried out in a wavelength-by-wavelength fashion, that is, we first apply the PCA to each of the IRDIS H2 and IRDIS H3 filters, and then co-add the images after PCA analysis. Fig.~\ref{fig:sphere_irdis_images} (left column) shows the final IRDIS images of the disk, as processed with GRAPHIC/PCA.

We followed a similar data reduction approach for the SPHERE/IFS frames, with a few key differences. We used the \texttt{vlt-sphere} package \citep{vigan2015sirius,vigan2020ascl} to perform pre-processing, i.e. to turn each frame from a 2D image of spectra to a 3D spectral cube. This routine makes use of several esorex functions \citep{pavlov2008} and applies dark and flat corrections, calculates the positions of spectra in the 2D image and calibrates their wavelength, and creates an IFU flat. The routine also corrects bad pixels and detector cross-talk \citep[see][for details]{vigan2015sirius}. We then applied the GRAPHIC PCA routine as described above. We considered each wavelength separately during the PCA analysis to create a cube of 39 ADI-reduced images. We also co-added the cube of PCA-processed images to create a broadband YJ reduction. We were unable to identify any disk signal in the reduced IFS data, even when all of the wavelengths were stacked and combined. The processed IFS images are shown in Fig.~\ref{fig:sphere_ifs_images}.

\subsection{ADI with TLOCI and ANDROMEDA}
\label{sec:spheredc}

To further rule out the possibility of the disk signal being a reduction artefact, we also reduced the data from scratch at the SPHERE Data Center (DC) by means of the SpeCal pipeline \citep{delorme2017,galicher2018}. 

For IRDIS data, we applied ADI using the standard Template Locally
Optimized Combination of Images (TLOCI) reduction technique
\citep{lafreniere2007}. Similar to the PCA approach, this involves building a custom reference image to subtract stellar speckles from each science observation. The calibration of true north and pixel scale employs observations of compact stellar clusters as described in \citet{maire2016}. These reductions are shown in the middle column of Fig \ref{fig:sphere_irdis_images}.

For the IFS data, we used the ANDROMEDA reduction technique \citep{cantalloube2015}. This tracks planetary signatures by using a forward-modeling approach and a maximum likelihood estimator \citep{mugnier2009}. By using the ADI technique and performing pairwise subtraction of the images, the quasi-static stellar halo is removed, while the circumstellar signal rotates through the observation sequence as a function of parallactic angles. The ANDROMEDA reduction of the IFS data is shown in Fig.~\ref{fig:sphere_ifs_images}.

The SPHERE DC results for both the IFS and IRDIS support our results with GRAPHIC: the disk is clearly visible in all IRDIS images except for 2016-03-16, and is not visible in any IFS reductions. We also searched for additional companions in the IFS field of view. While a number of points with SNR$\sim$4 are observed in both the GRAPHIC and ANDROMEDA IFS reductions, these are not consistent between epoch or between reduction method, implying that these are artefacts of the reduction process, as opposed to bona fide companions.

\subsection{RDI with reference library and PCA}
\label{sec:rdi}

Different techniques developed to remove the stellar contributions may have different impacts on circumstellar objects, especially extended objects (e.g., debris disks), for which the photometry and the morphology are affected \citep{milli2012}. To validate the disk detection via GRAPHIC/PCA and DC/TLOCI, we applied reference-star differential imaging \citep[RDI;][]{smith1984,golimowski2006} to recover the scattered light emission of the disk. 

We used the reference library implementation from \citet{xie2022} to construct and subtract reference images; we briefly describe the key steps here and refer the reader to that work for details. In this approach, a library of archival reference star observations are used to create the model of the stellar PSF. We adopted the archived observations of reference stars selected by \cite{xie2022}. The selected master reference library contains 725 SPHERE observations of reference stars in the DB\_H23 IRDIS filter. We used the mean square error (MSE) between images to measure the image correlation, and down-selected the 500 most correlated reference images from the master reference library for each science image in each epoch of science observations. For each science observation, we combined the down-selected reference images and formed a single library with non-redundant references. Then we performed PCA on the down-selected reference library to construct the PSF model for each science image in each epoch of HD\,112810 observation. For RDI-PCA, we used 400 PCA components to remove the stellar contributions. The temporal images were then derotated and mean combined to form the residual images in H2 and H3. The H2 and H3 bands were processed separately and mean combined to form the final residual image in the H23 band shown in Fig.~\ref{fig:sphere_irdis_images} (right column). 
The reference library of \citet{xie2022} does not include BB\_H data, since there are many fewer BB\_H observations that DB\_H23 observations in the SPHERE archive. We therefore only used the RDI approach for the 2016 data, observed in DB\_H23.

\section{Constraints on the Disk Structure}
\label{sec:model}

We modelled the scattered light emission of the HD\,112810 disk in order to constrain the disk structure and understand the physical distribution of small dust grains in the disk. In this section we first describe the modelling process, and then detail the various physical models that we tested. We found HD\,112810 to be well-fit by a single ring of dust at \diskrad, that is radially very narrow and vertically unconstrained. We used the epoch (6) observations collected with the IRDIS B\_H filter for the modelling, since this is the deepest disk observation and the highest signal-to-noise disk detection. We briefly discuss the best fit for other epochs in Sect.~\ref{sec:model-other-epochs}.

We used the same disk modelling procedure as in \citet{matthews2017} (see also \citealt{wahhaj2014}): our process consisted of generating synthetic disk images, subtracting these from the cleaned data, reprocessing that subtracted data, and measuring the goodness of the fit in the reprocessed, residual images. This allows for the modelling to take into account the radially varying transmission of the speckle removal process, as well as any self-subtraction introduced during the speckle removal.

\subsection{Synthetic disk model}

Our synthetic disk images were generated using the GRaTeR ray tracing code \citep[][IDL implementation]{augereau1999}. We modelled the disk as optically thin, meaning that the disk brightness at each point in 3D space is simply defined by the local dust density, the scattering properties of the dust, and the distance of that point from the star. 

The generalized form of the disk density in GRaTeR is 
\begin{equation}
\label{eq:gratermain}
    \rho \propto \frac{ \exp  \left( \left[\frac{-|z|}{\xi_0}\left(\frac{r}{r_0}\right)^{-\beta}\right]^\gamma \right) }
    {\sqrt{\left(\frac{r}{r_0}\right)^{-2\alpha_\mathrm{in}}+
        \left(\frac{r}{r_0}\right)^{-2\alpha_\mathrm{out}}}}
\end{equation}
with $r$ and $z$ the radial and vertical distance from the star respectively. The vertical structure is parametrized in the exponential term by $\xi_0$, $\beta$ and $\gamma$: $\xi_0$ is the disk vertical scale height at $r_0$, $\beta$ is the disk flaring index, and $\gamma$ defines the shape of the vertical profile. For this work, we set $\beta$=1 and $\gamma$=2 throughout -- that is, a disk with linear flaring and a Gaussian vertical profile.

The radial structure is set in the denominator by a characteristic radius $r_0$ and two power laws defined by $\alpha_\mathrm{in}$ and $\alpha_\mathrm{out}$, characterising how sharply the dust density decreases with distance from $r_0$. In the case that $|\alpha_\mathrm{in}| = |\alpha_\mathrm{out}|$, the radial maximum dust density is at radius $r_0$. Note that the quoted $r_0$ values and uncertainties refer to the characteristic radius, while the belt width is defined by the $\alpha_\mathrm{in}$ and $\alpha_\mathrm{out}$ parameters.

The scattering is parameterized as a function of angle using the Henyey-Greenstein function: 
\begin{equation}
p(\theta) = \frac{1}{4\pi} \frac{1-g^2}{(1-2g\cos\theta+g^2)^{3/2}}
\end{equation}
where $\theta$ is the scattering angle at each point in the disk. $g$ is the anisotropy parameter, with $g$=-1 and $g$=1 corresponding to the extreme cases of 100\% back-scattering and 100\% forward-scattering.

The disk flux $F$ at each point in the grid is then derived by combining the local disk density, scattering phase function, and the distance from the star ($d^2=r^2+z^2$) as
\begin{equation}
F \propto \frac{\rho \times p(\theta)}{d^2}
\label{eq:graterflux}
\end{equation}
Finally, the synthetic disk image is generated by adding the flux terms along the line of sight. For all fits presented in this paper, we used an arbitrary scaling factor of $\rho_0$ when subtracting synthetic disk images from the data.

\subsection{Parameter estimation process}

Synthetic images were generated following the process described above for each set of physical disk parameters, and were then convolved with a 2D-Gaussian to mimic the impact of the telescope PSF. These disk images were subtracted from each slice of the pre-cleaned data cubes (see Section \ref{sec:dr} for details), at the appropriate rotation angle for that slice. This subtraction of disk images was performed before the application of the stellar speckle removal algorithm. We used the GRAPHIC/PCA (Sect.~\ref{sec:dr-graphic}) speckle analysis process for model-subtracted cubes, to generate a residual image. PCA can be a relatively time-intensive procedure, and so we made two changes relative to our original GRAPHIC/PCA reduction described in Sect.~\ref{sec:dr-graphic}. Firstly, we only performed PCA on the central region of the image, where there is disk flux. We restricted the analysis to a radius of 91 pixels (=1.115\arcsec) from the central star. Secondly, we binned temporally to reduce the number of independent images in the datacube. We median-combined each consecutive set of 4 images, accounting for the small parallactic angle changes (always $<$1$^\circ$, corresponding to $<$1.2pixel at the disk ansae) between individual images.

This residual image was then used to calculate a $\chi^2$ value, as a proxy for the quality of the fit. This $\chi^2$ was calculated as $\sum{(F_i^2/\sigma_r^2)}$ with $F_i$ the data value of the pixel in the residual image, and $\sigma_r$ a radially varying uncertainty term, so as to represent the increased stellar speckle noise very close to the star. The $\sigma_r$ value is calculated from the standard deviation of pixel values in a small annulus, and inflated to account for the spatial correlation of neighboring pixels. Following \citet{hinkley2021}, we inflated the errors by the square root of the number of pixels in a correlated region, taking into account both the size of the PSF and the radial PSF elongation that derives from the filter bandwidth (which is negligible for the DB\_H23 narrowband filters). The inflated uncertainties are calculated as $\sigma(r)  = \sigma_{pixel}(r) \times \sqrt{FWHM \times r\Delta\lambda/\lambda} $, where $\sigma_{pixel}(r)$ represents the uninflated pixel uncertainty, calculated as the standard deviation of pixel values in each annulus. 

The $\chi^2$ was calculated in a bar-shaped region around the approximate position of the disk.
The $\chi^2$ value for each synthetic image allows us to assess the relative quality of each fit, and to optimize over disk parameters. We first found a visual fit that appeared to match the observed disk structure, and then performed an MCMC using \texttt{emcee} \citep{foremanmackey2013}, and using the visual fit values as the starting parameters.

\renewcommand{\arraystretch}{1.5}
\begin{table*}[t]
    \centering
    \caption{MCMC best fit model parameters for each of the fits described in Sects.~\ref{sec:5parfit}-\ref{sec:diskradial}.}
    \footnotesize
    \begin{tabular}{l|c|c|c|l}

    Parameter & 5-parameter fit & Vertical profile fit & Radial profile fit & Prior \\
    \hline \hline
$i_\mathrm{tilt}[^{\circ}]$ & $71.5^{+2.0}_{-2.1}$ & $72.6^{+1.9}_{-2.0}$ & $75.7^{+1.1}_{-1.3}$ & [60, 88], uniform in cos(i)  \\
$PA[^{\circ}]$ & $98.6^{+0.8}_{-0.9}$ & $98.6\pm0.9$ & $98.6\pm0.7$ & [91, 107], uniform  \\
$r_0[au]$ & $101\pm7$ & $104\pm7$ & $118\pm9$ & [50, 180], log uniform  \\
log10($\rho_{0}$) & $8.44\pm0.10$ & $8.88^{+0.51}_{-0.36}$ & $8.32^{+0.08}_{-0.07}$ & [1e7, 1e12], log uniform \\
$g$ & $0.49^{+0.12}_{-0.11}$ & $0.48^{+0.13}_{-0.11}$ & $0.65^{+0.09}_{-0.10}$ & [-1, 1], uniform \\
$\xi_{0}[au]$ & 4.4 (fixed) & $1.3^{+2.0}_{-0.9}$~~\tablefootmark{(a)} & 4.4 (fixed) & [0.1, 50], uniform \\
$\alpha_{in}$ & 3.5 (fixed) & 3.5 (fixed) & $11^{+8}_{-5}$ & [0, 25], log uniform \\
$\alpha_{out}$ & -3.5 (fixed) & -3.5 (fixed) & $-9^{+3}_{-5}$ & [0, 25], log uniform \\
    \end{tabular}
 \tablefoot{For each parameter the median and 68\% confidence interval of the posterior are quoted.$\rho_0$is an arbitrary scaling factor; PA is measured anti-clockwise of North. Note that $r_0$ is the characteristic radius, with the disk width being defined by $\alpha_{in}$ and $\alpha_{out}$.  
 \tablefoottext{a}{The $\xi_{0}$ posterior closely follows the prior distribution, and should not be taken to represent the true scale height of the disk.}
    }
    \label{tab:disk_parameters}
\end{table*}
\renewcommand{\arraystretch}{0.666}

\subsection{Five parameter fit}
\label{sec:5parfit}

We first tested a simple model with five free parameters: three disk parameters (radius r$_{0}$, forward scattering $g$, and overall scaling $\rho_0$), and two parameters defining the on-sky position of the disk (inclination $i_\mathrm{tilt}$, and position angle $PA$). For this fit, we set $\xi_0$=4.4au. This provides an aspect ratio of $\sim$4\% at $r_0$ (initially estimated at $\sim$110au), which is the minimum natural aspect ratio of small dust grains in a disk, in the absence of perturbing bodies \citep{thebault2009}, and we set $|\alpha_\mathrm{in}|=|\alpha_\mathrm{out}|=3.5$.

\begin{figure*}[t]
    \centering
    \includegraphics[width=\textwidth]{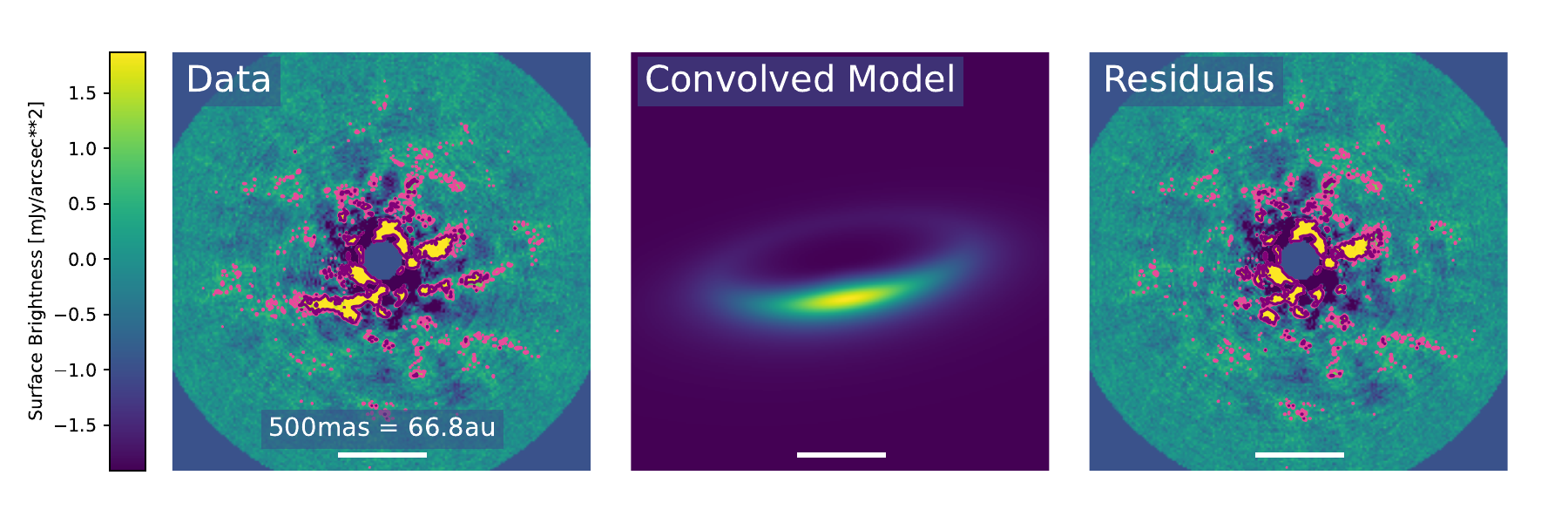}
    (a) 5-parameter fit
    \includegraphics[width=\textwidth]{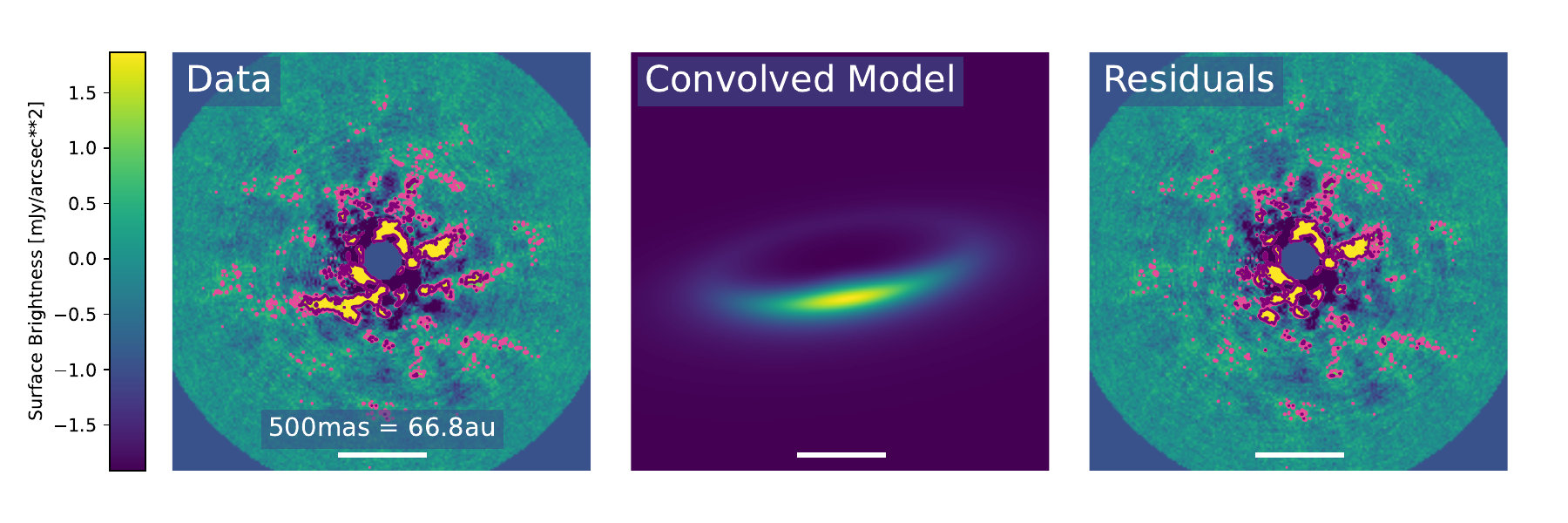}
    (b) vertical structure fit
    \includegraphics[width=\textwidth]{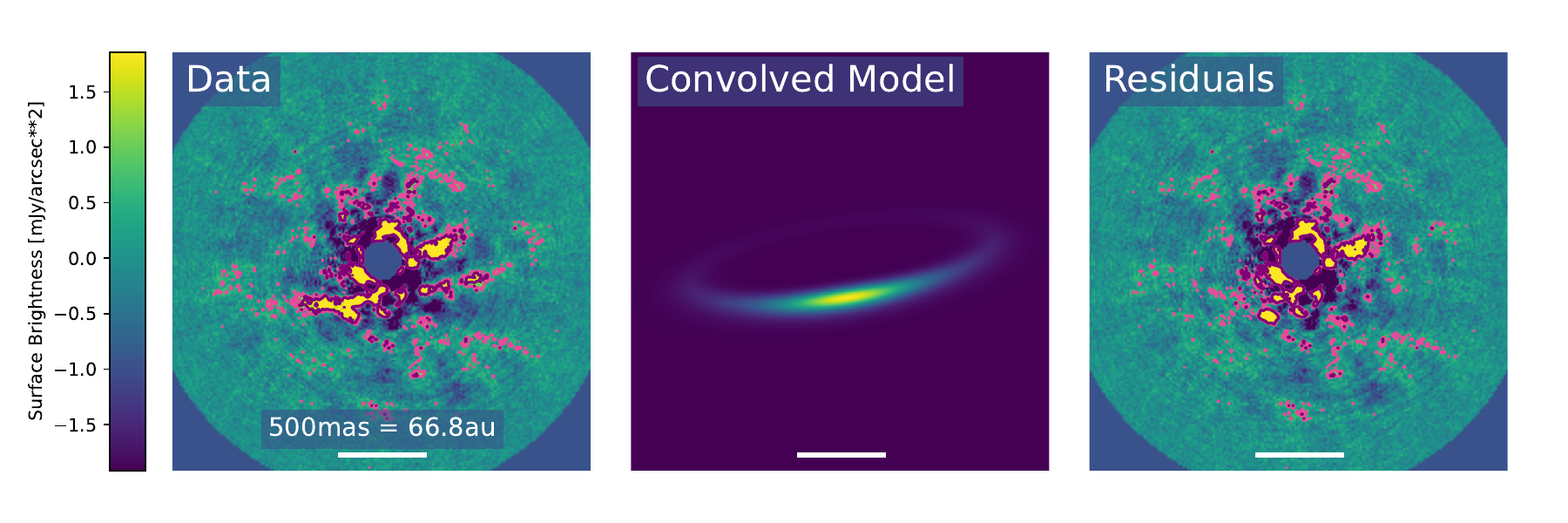}  (c) radial structure fit
    \caption{Best fitting models to the data, alongside the raw data and the residual image with the disk subtracted. 3- and 5-$\sigma$ contours (relative to the background noise) are shown in pink and purple respectively. The three rows show the best fitting models for (a) the 5-parameter fit (b) the vertical structure fit, and (c) the radial structure fit (see Sects.~\ref{sec:5parfit}--\ref{sec:diskradial}). Best-fit parameters are given in Table~\ref{tab:disk_parameters}. Residual images are generated by subtracting model disks from the raw data, and then repeating the PCA reduction. The debris disk is well-subtracted in all cases.}
    \label{fig:gratermodel}
\end{figure*}

We found a good fit to the disk, which is shown in Fig.~\ref{fig:gratermodel} (top panel), alongside the cleaned data without the disk subtracted, and a residual image (that is, a cleaned datacube with the best-fit model subtracted that is then re-processed with PCA). No significant structure seen in the residuals, indicating that the disk is well-fit with a single dust belt at radius $101\pm7$au. The best-fitting parameters are listed in Table \ref{tab:disk_parameters}. In Appendix \ref{app:extramodelling} we include a corner plot of the posterior, highlighting the key correlations in this dataset. In particular, $r_0$ and $i_\mathrm{tilt}$ are correlated, which is a natural result of the disk geometry: the observable parameter from which the $r_0$ and $i_\mathrm{tilt}$ values are derived is the semi-minor axis of the projected disk ellipse in the SPHERE images, which is equal $r_0 \cos (i_\mathrm{tilt}) $.

A bright patch of emission is seen to the right of the star, which is not co-located with the disk emission; this feature is only seen in one data epoch  and is assumed to be an uncorrected stellar speckle. 

\subsection{The disk vertical profile}
\label{sec:diskvertical}

To test whether the disk vertical profile can be usefully constrained with the in-hand data, we also performed a fit with one additional free parameter, the disk scale height $\xi_0$, in addition to the five parameters explored in Sect.~\ref{sec:5parfit}. We used a log-uniform prior for $\xi_0$, and used best-fit values from the 5-parameter fit as starting values for all other parameters.

The best fit for this model is shown in Fig.~\ref{fig:gratermodel} (second row). The posterior for $\xi_0$ traces the prior distribution. The other disk parameters are in good agreement with the values found in Sect.~\ref{sec:5parfit}, except that $\rho_0$ is somewhat higher and less well-constrained, due to the strong anti-correlation between $\rho_0$ and $\xi_0$. A corner plot of the posterior distribution is included in Appendix \ref{app:extramodelling}.

While formal median and best-fit values are derived from the posterior and quoted in Table \ref{tab:disk_parameters}, we note that these are largely driven by the prior, and in the absence of data constraints a log-uniform prior favors small values. The median and best-fit are lower than the physically motivated natural aspect ratio used in Sect.~\ref{sec:5parfit}; we infer that the vertical structure cannot be constrained, and therefore prefer the physically motivated value of $\xi_0$=4.4.

\subsection{The disk radial profile}
\label{sec:diskradial}

We also explored whether the disk radial profile can be usefully constrained with the available data. In this case we performed a fit with seven free parameters: the five parameters from Sect.~\ref{sec:5parfit}, as well as $\alpha_\mathrm{in}$ and $\alpha_\mathrm{out}$; we fixed the $\xi_0$=4.4 as discussed in Sect.~\ref{sec:diskvertical}. We used log-uniform priors for both $\alpha_\mathrm{in}$ and $\alpha_\mathrm{out}$.

The best fit for this model is shown in Fig.~\ref{fig:gratermodel} (bottom row). We find evidence that the disk is narrow: for both $\alpha$ parameters, low values are disfavored. In particular, values of $\alpha_\mathrm{in}\lesssim$7.5 are disfavored; above this value the posterior follows the prior. For $\alpha_\mathrm{out}$, the posterior is moderately well constrained, though very high values are not completely excluded; we find a best-fit value of $\alpha_\mathrm{out} = 8.9^{+4.9}_{-3.2}$. We interpret that the scattered-light disk is narrow -- broad disks are ruled out -- and that the disk is too narrow to be well-resolved, precluding better constraints on the radial structure to be derived. Effectively, this fit is providing a lower limit on the $\alpha$ parameters, and an upper limit on the disk radial width.

This 7-parameter fit gives a somewhat higher radius than the 5-parameter fit ($118\pm9$au vs $104\pm7$au). This is a natural result of the disk geometry: while the dust density is at a maximum at $r_0$, the brightness profile is proportional to $r/d^2$, and is therefore at maximum closer to the star that $r_0$ -- a broad disk profile characterized by $|\alpha_\mathrm{in}|=|\alpha_\mathrm{out}|=3.5$ has maximal brightness at $0.91r_0$, a very narrow disk profile with $|\alpha_\mathrm{in}|=|\alpha_\mathrm{out}|=10$ has maximal brightness at $0.99r_0$. The offset between observed maximal brightness and $r_0$ is much smaller for this seven parameter fit with a very narrow disk profile than for the fit presented in Sec.~\ref{sec:5parfit}.

Due to the correlation of $r_0$ and $i_\mathrm{tilt}$, we also retrieve a somewhat higher inclination value than the 5-parameter fit (75.7$^{+1.1\circ}_{-1.3}$ vs 72.6$^{+1.9\circ}_{-2.0}$ ). Both $r_0$ and $i_\mathrm{tilt}$ are $\sim$1.4$\sigma$ larger for the 7-parameter fit than the 5-parameter fit. 

We adopt the results of this 7-parameter fit as the derived disk parameters. The scattered light images of HD\,112810 hosts a single, radially narrow ring, with a poorly constrained vertical profile.

\subsection{Disk constraints from other epochs}
\label{sec:model-other-epochs}

All the above sections have focused solely on the 2020-02-18 data (observation 6 in Table \ref{tab:obslog}), which is the highest significance disk detection due to the long exposure time and broadband filter. We also explored the constraints available from other epochs of data. The 2020-02-18 observations are the most constraining: in particular, the radial profile (that is, $\alpha_\mathrm{in}$ and $\alpha_\mathrm{out}$) cannot be well constrained from other epochs of data, which also causes small differences between epochs in the retrieved $r_{0}$ and $i_\mathrm{tilt}$ as discussed in Sec.~\ref{sec:diskradial}.

\section{Limits on Substellar Companions}
\label{sec:planetlimits}

Stellar and substellar companions to HD\,112810 can be observationally excluded across much of parameter space. As well as the high-contrast imaging observations presented in this work, there are archival radial velocity data, and on-sky acceleration constraints based on the \hipp~and \gaia~observations of the systems.

\subsection{Observational limits on planets: imaging}

\begin{figure*}[t]
    \centering
    \includegraphics[width=\textwidth]{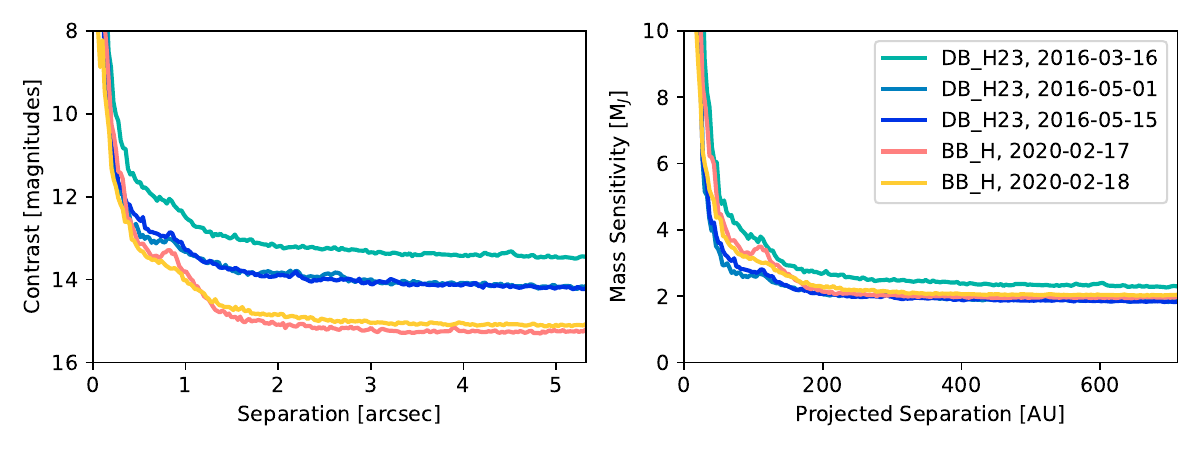}
    \caption{Contrast (left) and mass sensitivities (right) of each epoch of observations. The legend applies to both figures; cool colours indicate the narrowband DB\_H23 observations and warm colours indicate the broadband BB\_H observations. Companion magnitude limits were converted to mass limits using the ATMO models from \citet{phillips2020}.} 
    \label{fig:contrastcurves}
\end{figure*}

To calculate the sensitivity of the SPHERE images, we injected fake companions into the cleaned data cubes, before the processing with the PCA algorithm. We then repeated the full PCA reduction, and measured the brightness of the fake companions in the PCA reduced cubes. This process was repeated a number of times, with fake planets at different position angles and separations. Fig.~\ref{fig:contrastcurves}(left) shows the sensitivity of our images in magnitudes relative to the host star. To determine our ability to detect planets in this dataset, we additionally converted these magnitude values to masses. Absolute masses were converted to contrasts using the ATMO isochrones from \citet{phillips2020} in the relevant SPHERE filters using an age of 17Myr. The mass sensitivities in the IRDIS bands are given in Fig.~\ref{fig:contrastcurves}(right).

\label{sec:detlim-hci}
We also converted these limits to completeness curves, using \texttt{exoDMC} \citep{exodmc}, as shown in Figure \ref{fig:det_lim_combined}. \texttt{exoDMC} simulates a suite of test planets, and calculates the fraction of planets that would be observed as a function of mass and separation. For this process, we fixed the inclination of all test planets to the inclination of the disk.

The young age of this system corresponds to an excellent planet sensitivity in the background limited regime, though this regime is limited to $\gtrsim$150au due to the distance of the star (133.66$\pm$0.29pc). We are sensitive to planetary mass companions ($<$13M$_\textrm{J}$) beyond $\sim$150mas=21au, and to $\sim$2M$_\textrm{J}$ companions in the background limited regime (beyond $\sim$200au).

\subsubsection{Candidate companions}

Five candidate companions to HD\,112810 were identified in \citet{matthews2018}, with separations between 3.5\arcsec and 6.2\arcsec from the host star. In that work, HD\,112810 was only observed once, and the candidate companions were deemed to be likely background objects due to their relatively wide projected separations (between 470au and 830au). Here, we use the new data from 2020 to reassess the status of these candidate companions, and we confirm that they are indeed background objects (full astrometry and plots are given in Appendix \ref{app:cc_astrometry}). We also note that two of these sources are detected in \gaia~DR3; the \gaia~detections are consistent with the results of the common proper motion test.

We also present astrometry and photometry for one additional candidate companion identified in the 2020 data. This object was too faint to be detected in the 2016 observations (see Fig.~\ref{fig:contrastcurves} for relative sensitivities of the different observations), but is observed in both 2020 epochs. At separation $\sim$5.2\arcsec, this candidate is highly likely to be a background object given the local background density, and the wide separation of the companion. Full astrometry for this companion is given in Appendix~\ref{app:cc_astrometry}.

\subsection{Observational limits on planets: radial velocities}
\label{sec:detlim-rv}

HD\,112810 was observed with HARPS as part of a larger radial velocity survey for massive planets around young stars \citep{grandjean2023}. That work collected 16 radial velocity points across 8 nights between 2019 Apr 29 and 2021 Feb 28. We independently reduced these data following the process described in \citet{trifonov2020}, and then tested the sensitivity of these data to substellar companions by injecting synthetic companion signals across a range of masses and periods, at the inclination of the debris disk. This process was carried out using an updated version of the code from \citet{barbato2018}.

\begin{figure*}[t]
    \centering
    \includegraphics[width=1.\textwidth]{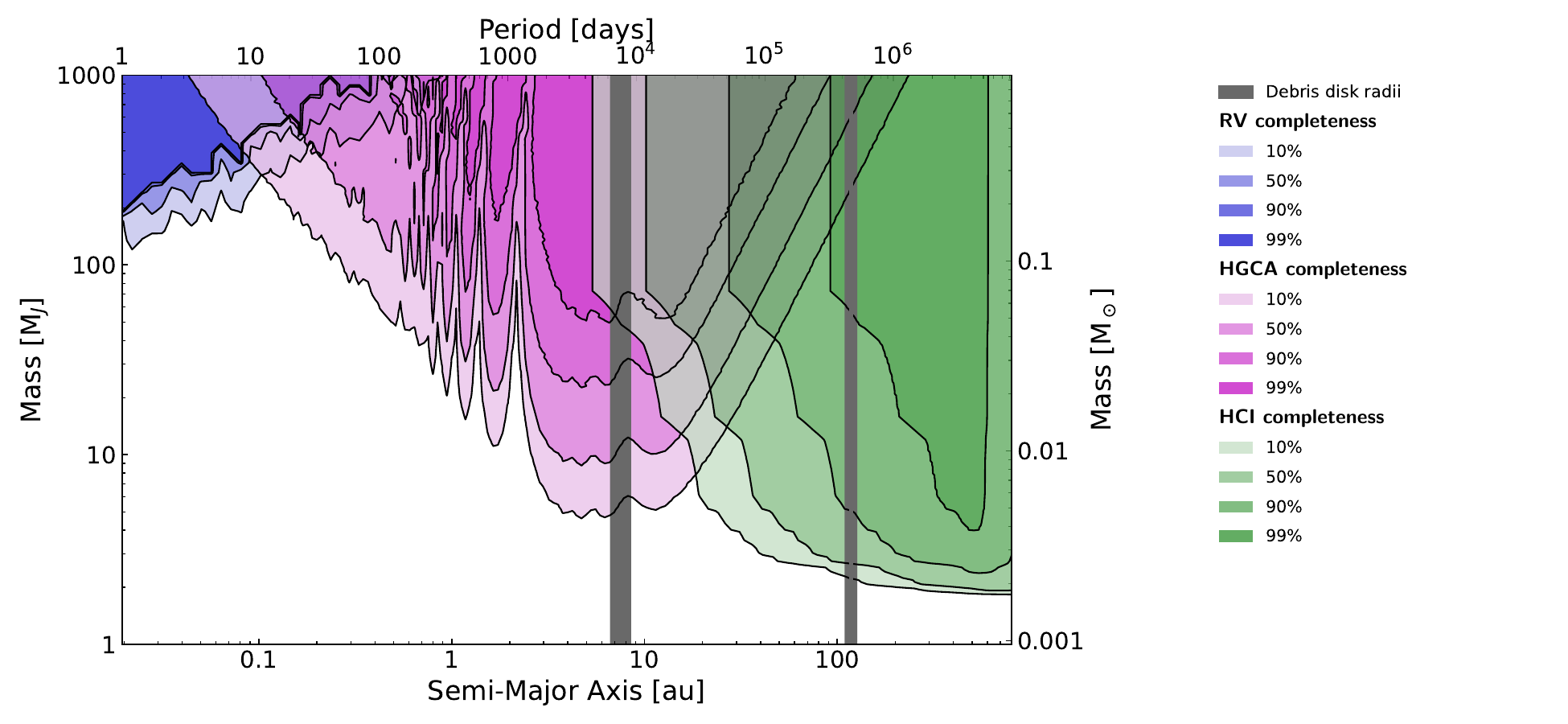}
    \caption{Detection completeness for HD\,112810 derived from radial velocities (blue) and \hipp-\gaia~proper motion analysis (pink), and high contrast imaging (green) along with the debris belt positions (grey). Shaded regions indicate the mass/separation threshold at which 10\%, 50\%, 90\% and 99\% of companions would be detectable with in-hand data, as described in Sects.~\ref{sec:detlim-rv}, \ref{sec:detlim-hgca} and \ref{sec:detlim-hci} respectively. The inner and outer belt positions are derived from the SED fit and SPHERE modelling respectively. Planetary mass companions and small brown dwarfs at intermediate and wide separations are largely ruled out from the proper motion and imaging constraints. The radial velocity constraints are weaker, ruling out only massive stellar companions at short periods.}
    \label{fig:det_lim_combined}
\end{figure*}

We found that the radial velocity data for this target are not sufficient to place constraints on substellar companions. This is perhaps unsurprising, given that the host star is a young and early-type star: young stars are typically active, and early-type stars have fewer lines, making this a very challenging host star for radial velocity planet detection. The data are only sensitive to companions $\gtrsim$200M$_J$(=0.19M$_\odot$) at periods of a few days, and even more massive companions at wider separations (See Figure~\ref{fig:det_lim_combined}).

\subsection{Observational limits on planets: astrometry}
\label{sec:detlim-hgca}

We also examined the constraints available from astrometric analysis of HD\,112810, based on the \hipp~and \gaia~proper motion measurements: a significant difference in proper motion would indicate that a target is accelerating, due to the presence of an unseen companion. HD\,112810 does not show any sign of acceleration: the short-term proper motions of HD\,112810 at the \gaia~epoch and \hipp~epoch, as well as the long-term average proper motion (calculated from the \hipp~and \gaia~position measurements) are all consistent with each other \citep{brandt2021}. The star has a Renormalized Unit Weight Error (RUWE) value of 0.994 \citep{lindegren2018,gaia_dr3} -- strongly implying that the object is a single star. To determine the companion space that is ruled out by these observations, we simulated the observed proper motions for synthetic planets across a range of masses and periods, at the inclination of the debris disk and randomizing over orbital eccentricity, and compared their proper motions to the measured \hipp, \gaia, and long-term average proper motion. We then calculated the fraction of simulated orbits that are incompatible with the observational constraints. Figure~\ref{fig:det_lim_combined} (pink shading) indicates the threshold at which the in-hand data rules out 10\%, 50\%, 90\% and 99\% of all simulated planets. Stellar companions are largely ruled out beyond $\sim$4au; our detection completeness to brown dwarfs ($>$13M$_J$) is above 50\% beyond 3au based on the proper motion of HD\,112810.

\subsection{Planet inferences from disk properties}
\label{sec:dynamicalplanetlimits}

Inferring the presence of planets using the disk structure -- either from resolved images or SED constraints -- is a method that has long been employed since the discovery of debris disks \citep[see e.g.,][]{mouillet1997,quillen2006}. In particular, when two belts of observed with a gap in between, one common hypothesis is that one or more planets may have carved out this gap -- in this case, the inferred planet properties can be constrained from the disk structure \citep[see][for the most recent applications of this method dedicated to the disks of HD 107146, HD 92945, HD 15115, and HD 206896, respectively]{marino2018,marino2019,macgregor2019,marino2020}. General studies using both analytical arguments and N-body simulations -- and exploring the different processes through which planets can carve gaps in disks -- allow us to derive the possible number, mass and location of these planets for any given system \citep[e.g.][]{morrisonmalhotra2015,shannon2016,yelverton2018}.

The SED of HD\,112810 indicates the presence of two belts, located at $\sim$8au and $\sim$160au \citep{chen2014,pawellek2015} -- somewhat reminiscent of the HR\,8799 architecture, which has dust belts at $\sim$15au and $\sim$145au \citep{booth2016}. In the general framework provided by \citet{shannon2016}, if these two rings formed as a single broad belt with a gap carved by planets, that would imply a chain of at least 2 planets, with mass at least 1.8M$_\textrm{J}$, close to the planet detection limits of the current work (Fig.~\ref{fig:contrastcurves}). However, we note that \citet{shannon2016} considered only circular planets, and eccentric planets may have a similar gap-carving impact at smaller masses \citep[e.g.][]{bonnefoy2017}. Further, the properties of the inner debris disk belt are derived from the SED only, and are somewhat uncertain. These very widely spaced belts of debris could also indicate two rings of material that are generated separately, as opposed to a single belt with a gap carved by planets. More detailed disk constraints would be required to make detailed dynamical inferences about planets in the system - in particular, such inferences would benefit from observations with the spatial resolution and sensitivity to resolve the disk radial profile, as well as detect any fainter disk components that are not detected in the current images.

\section{Discussion}
\label{sec:discussion}

HD\,112810 provides a particularly faint scattered-light disk, barely detectable with present-day facilities: even with several long observations (1-2 hours on-sky integration per visit) with a state-of-the-art AO+coronagraph system, the disk is only weakly detected. This is perhaps unsurprising, given its large radius ($\sim$118au), moderate inclination ($\sim$75.7$^\circ$), and the relatively low infrared excess of the outer belt (\Lir=5.9$\times10^{-4}$) placing the debris disk at the edge of the population of disks that have been imaged in scattered light \citep{esposito2020}.

This comparative faintness inevitably makes the disk challenging to study. The disk is nonetheless a valuable addition to the gallery of resolved disks, in particular due to its moderate inclination angle, which traces out a similar geometry as HR\,4796.

\subsection{The GRATER disk model: a single, symmetric ring}

In Sect.~\ref{sec:model}, we demonstrated that the debris disk is well fit as a single ring, with a characteristic radius \diskrad, a poorly constrained vertical profile, and a very narrow radial profile (see Sect.~\ref{sec:model}; Fig.~\ref{fig:gratermodel}).

We highlight that some disk parameters are correlated; in particular the radial profile, the characteristic radius $r_0$, and the disk inclination angle. That is, without an accurate determination of the radial profile, the true characteristic radius $r_0$ will be systematically offset from the derived value; this in turn has an impact on the derived inclination angle of the disk. This effect would particularly impact comparative studies where fitted $r_0$ values from various different literature sources are compared, depending on the radial profile assumptions made. Ideally, such comparative works should rederive disk parameters using consistent modelling assumptions.

We do not see any evidence for complex structure in the HD\,112810 disk (i.e.~multiple rings, gaps, warps, asymmetries etc). However, it is worth noting that this is partly a comment on the relative faintness of the disk. For example, a secondary ring of small grains, such as that observed for HIP\,73145 \citep{feldt2017} and HIP\,67497 \citep{bonnefoy2017}, would likely be fainter than the main component observed here, and below the noise floor of these observations. 

\subsection{Comparison of scattered light, thermal emission, and disk SED}

Near-IR scattered light images, such as those presented here, trace a different population of grains than those that dominate the ALMA images and SED measurements for the system. We nonetheless briefly comment on how the parameters compare between the SPHERE, ALMA and SED observations. 

\citet{liemansifry2016} imaged HD\,112810 using ALMA, and presented both continuum at 1240$\mu$m and $^{12}$CO(2-1) images of HD\,112810. The disk is only detected in the continuum emission; in these images the disk is resolved only along the major axis. \citet{liemansifry2016} fit the debris disk as a single, broad belt, at an inclination angle $>$67$\deg$, and a position angle of 96$^{+15}_{-17}$$^\circ$. That work - updated with the \gaia~distance to the target - found the disk inner and outer edges to be R$_\textrm{in}<28$au and R$_\textrm{out}=130^{+80}_{-70}$au respectively.

The scattered light disk position angle and inclination derived in this work are in good agreement with the ALMA values. Notably, the disk is only marginally more inclined than the maximum resolvable inclination of $\sim$67$\deg$ presented in \citet{liemansifry2016}; a deeper observation with a longer baseline would likely be able to resolve both the major and minor axes of the disk. We measured a disk radius of \diskrad, which is consistent with the outer edge of the broad dust belt observed with ALMA for this target. The scattered light detection indicates a much narrower belt than the ALMA constraint, but we note that neither the scattered light nor the thermal imaging place particularly strong constraints on the dust radial profile.

Both the ALMA and SPHERE data indicate a disk at a smaller radius than predictions based on the SED alone. \citet{chen2014} used Spitzer IRS spectroscopy to determine that the star has a two-temperature infrared excess, which is well fit using two distinct dust temperatures of 258$^{+5}_{-7}$K and 57$^{+6}_{-4}$K. These temperatures can be used to infer orbital separations, following the scaling laws from \citet{pawellek2015}, which are derived for a variety of realistic grain compositions and dust particle porosities. These scaling laws suggest that the two temperatures correspond to rings of dust at $\sim$8au and $\sim$160au; the \citet{pawellek2015} prediction for the outer belt varies between 155au and 185au depending on the chosen dust composition, and has an uncertainty of $\sim$20au. In particular, the prediction for grains that are 50\% astrosilicate + 50\% vacuum is 155$^{+25}_{-22}$; our scattered light measurements are within $\sim2\sigma$ of this value. This small difference in SED and scattered light radius is consistent with the large scatter seen when comparing SED-derived radii with scattered light values: deriving radii from SEDs requires assumptions about the grain composition, grain size, and grain porosity -- all of which could vary between disks. A further complicating factor is that the small dust particles to which scattered light analysis is primarily sensitive do not exactly trace the location of the larger parent bodies, which dominate the SED emission.

\subsection{HD\,112810 compared to the population of scattered light disks}

\begin{figure*}[t]
    \centering
    \includegraphics[width=0.48\textwidth]{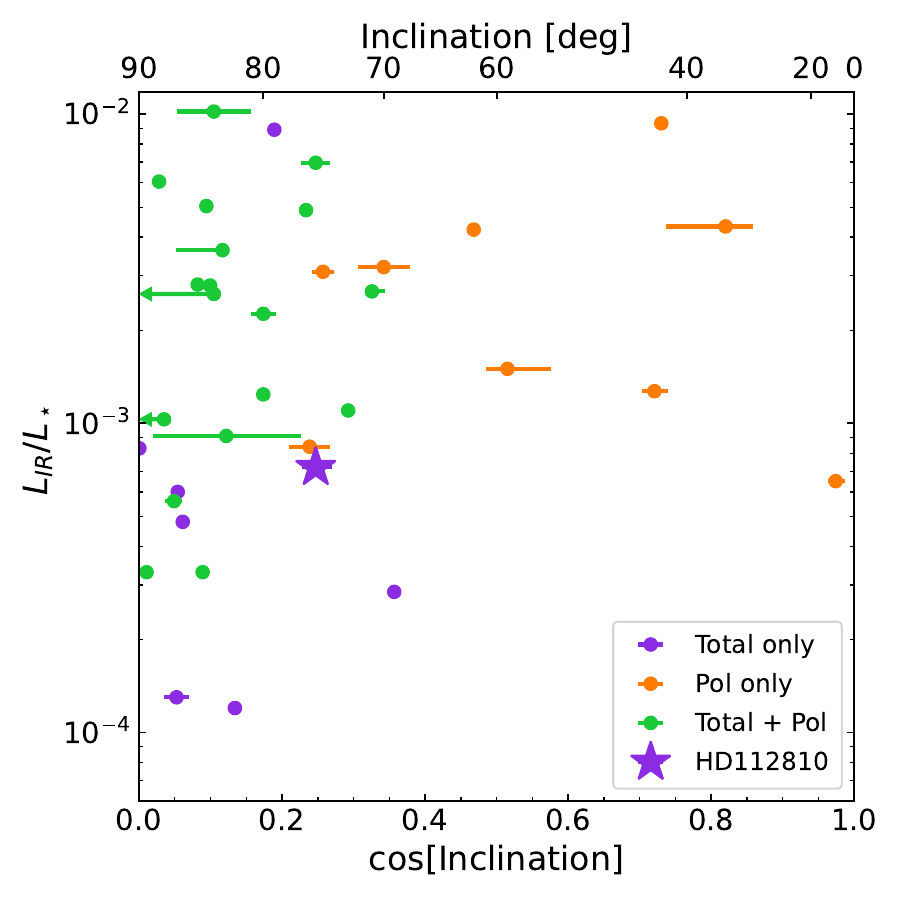}
    \includegraphics[width=0.48\textwidth]{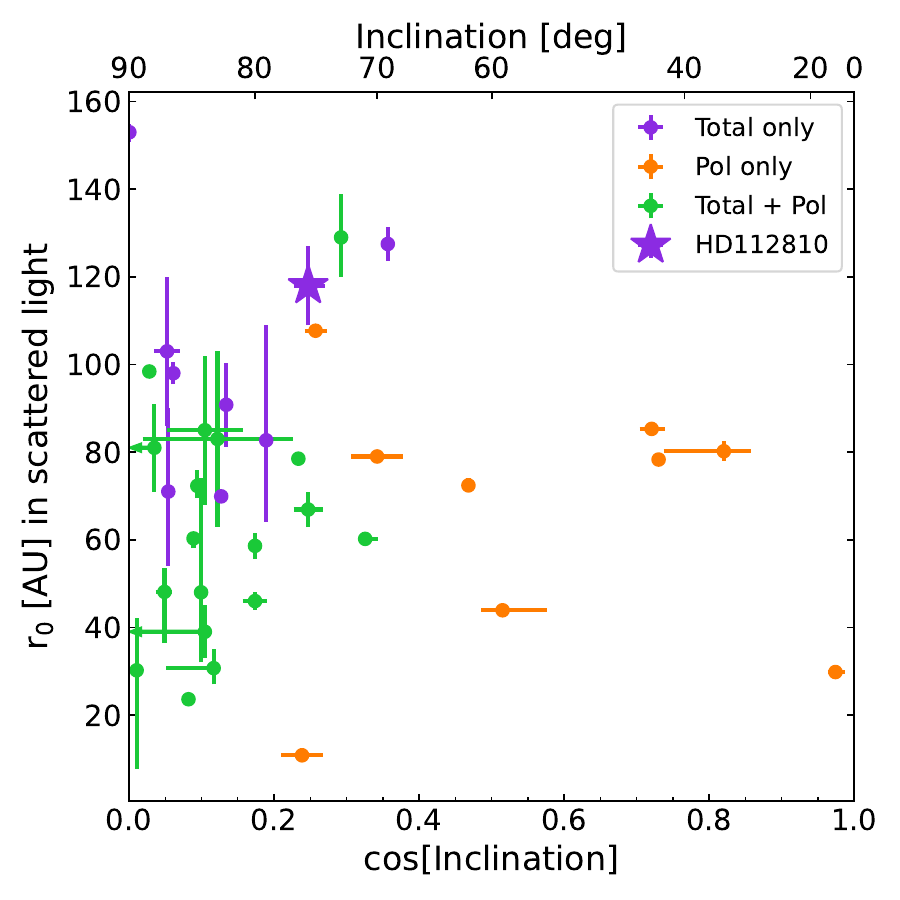}
    \caption{The HD\,112810 debris disk (indicated with a purple star) compared to the sample of debris disks that have previously been imaged from the ground. Disks are color coded based on detection method: orange for polarized light only, purple for total intensity only, and green for detection in both polarized and total intensity. \textit{Left:} fractional infrared excess and on-sky inclination of disks; HD\,112810 stands out as one of the least inclined disks imaged in total intensity. \textit{Right:} modelled peak dust density ($r_0$) and on-sky inclination of disks; in this case HD\,112810 stands out as one of the largest disks to be imaged in scattered light, and is strikingly similar in geometry to HD131835, as well as the slightly larger disks surrounding HD141011 and 49 Ceti. Literature disk properties are from the ensemble of objects presented in \citet{esposito2020} and from individual detections presented in \citet{choquet2017,bonnefoy2017,sissa2018,perrot2019,bonnefoy2021,hinkley2021,lawson2021,perrot2023,marshall2023}.}
    \label{fig:lir_vs_inc}
\end{figure*}

In this section we consider how HD\,112810 compares to the population of debris disks that have been detected in scattered light. Fig.~\ref{fig:lir_vs_inc} shows the disk inclination, $L_{IR}/L_\star$ and peak dust density of HD\,112810 as compared to the sample of disks that have previously been observed from the ground in scattered light.

Even with an inclination of \diskincl, HD\,112810 stands out as one of the least inclined disks to be detected in total intensity from the ground: only 49 Ceti (73$\pm$3$\deg$), HD\,117214 (71.0$^{+1.1}_{-0.4}$$\deg$) and HD141011 (69.1$\pm$0.9$\deg$) are less inclined \citep[][respectively]{choquet2017,esposito2020,bonnefoy2021}. The disk has a similar inclination to the heavily studied HR\,4796, as well as HD\,129590. The bias of ADI-based disk detection towards very edge-on disks is well-known, so this is unsurprising, yet disks slightly further from edge-on are some of the most interesting for studying the disk SPF. For very edge-on disks, the grains with the smallest and largest scattering angles are obscured by the stellar PSF; further from the star, the front and back edges of the disk may be blended due to the stellar PSF. The geometry of moderate-inclination disks, meanwhile, is such that almost the entire range of scattering angles is accessible, with no parts of the disk obscured or blended. In the case of HD\,112810, the SPF could be measured from $\sim$15$\deg$ to $\sim$165$\deg$. Such moderate-inclination disks are therefore particularly useful for gaining insight into the grain physics within the disk, since a broader range of scattering angles are accessible, especially if the back edge of the disk is detected. Indeed, HR\,4796 has proved particularly intriguing for SPF studies (see e.g. \citealt{milli2017} for a detailed discussion). HD\,112810 is sufficiently faint that such analyses is tricky with current data, but the disk is well-suited to SPF analysis with future facilities.

HD\,112810 also stands out as one of the largest debris disks imaged in scattered light, with a radius of $\sim118$au. The disk radius and inclination are strikingly similar to that of the Sco-Cen object HD\,131835, though that target has a much brighter fractional infrared luminosity than HD\,112810. As well as the polarimetric detection indicated in Fig.~\ref{fig:lir_vs_inc}, HD\,131835 was very recently detected in total intensity with SPHERE \citep{xie2022}. Unlike HD\,131835, HD\,112810 has not yet been detected in polarized light. Note however that we have only presented total intensity images in this work; given the geometry and the low \Lir of HD\,112810, a polarized light detection would be challenging: the disk has a lower \Lir than the majority of disks that have previously been detected in polarized light (Fig.~\ref{fig:lir_vs_inc}), with only very edge-on disks being detected at fainter \Lir values.

\section{Conclusions}

This paper has presented the first scattered light images of the HD\,112810 debris disk, which we observed several times with VLT/SPHERE. Our key findings are as follows:

   \begin{enumerate}
      \item We detected the HD\,112810 debris disk in scattered light, using H-band observations collected with the VLT/SPHERE/IRDIS high-contrast imaging platform. The signal is robust to our data processing approach, and is observed over several epochs of observations, but was not observed in the VLT/SPHERE/IFS observations in the YJ bands.
      \item This disk is best fit with a single-belt debris disk, with a radius \diskrad~and an inclination angle of \diskincl. The detection is consistent with a single, narrow ring, but the signal-to-noise ratio of the disk is insufficient to adequately constrain the radial/vertical profile of the dust. This geometry is consistent with partially resolved ALMA images of the debris disk. The disk appears moderately forward scattering, with a Henyey-Greenstein forward scattering parameter 0.6$\pm$0.1. 
      \item The disk lies towards the edge of the parameter space of disks that have been imaged in scattered light: it has a large radius, a moderate inclination, and a relatively low \Lir. This makes characterizing the disk challenging, but the disk geometry will allow for detailed study of the grain properties with future facilities.
      \item No bound companions were detected in this system with direct imaging, astrometry or radial velocity. Smaller planets that remain undetectable (i.e., less than a few M$_\textrm{J}$, depending on orbital parameters and phase) could nonetheless be responsible for carving the observed debris disks structure.
    \end{enumerate}

HD\,112810 adds to the growing inventory of scattered light disks, and extends the parameter space of the scattered light disk sample. The disk inclination of \diskincl could allow for the disk SPF to be measured between $\sim$15$\deg$-165$\deg$. The debris disk is potentially amenable to imaging with HST, which would probe the faintest surface brightness material further from the star, and with future facilities such as the ELTs.


\begin{acknowledgements}

We thank the anonymous referee for the timely and thorough referee report, which improved the quality of this manuscript.
\\
Based on observations collected at the European Organisation for Astronomical Research in the Southern Hemisphere under ESO programme 0101.C-0016(A), with additional data from programmes 096.C-0713(C), 097.C-0949(A) and 097.C-0060(A), 098.C-0739(A) and 1101.C-0557(D).
This work has made use of the the SPHERE Data Centre, jointly operated by OSUG/IPAG (Grenoble), PYTHEAS/LAM/CESAM (Marseille), OCA/Lagrange (Nice), Observatoire de Paris/LESIA (Paris), and Observatoire de Lyon.
We thank all the principal investigators and their collaborators who prepared and performed the observations with SPHERE. Without their efforts, we would not be able to build the master reference library to enable our RDI technique.
\\
This work has made use of data from the European Space Agency (ESA) mission
{\it Gaia} (\url{https://www.cosmos.esa.int/gaia}), processed by the {\it Gaia}
Data Processing and Analysis Consortium (DPAC,
\url{https://www.cosmos.esa.int/web/gaia/dpac/consortium}). Funding for the DPAC
has been provided by national institutions, in particular the institutions
participating in the {\it Gaia} Multilateral Agreement.
\\
This research has made use of the SIMBAD database, operated at CDS, Strasbourg, France.
\\

This work has been carried out in part within the framework of the NCCR PlanetS supported by the Swiss National Science Foundation under grants 51NF40\_182901 and 51NF40\_205606. E.C.M. and W.C. acknowledge the financial support of the SNSF.
C.D. acknowledges support from the European Research Council under the European Union’s Horizon 2020 research and innovation program under grant agreement No. 832428-Origins. 
This work has been supported by a grant from Labex OSUG (Investissements d’avenir – ANR10 LABX56). 
S.D. acknowledges support  has been supported by the PRIN-INAF 2019 "Planetary systems at young ages (PLATEA)" and ASI-INAF agreement n.2018-16-HH.0.  
This work is supported by the French National Research Agency in the framework of the Investissements d’Avenir program (ANR-15-IDEX-02), through the funding of the "Origin of Life" project of the Univ. Grenoble-Alpes.
B.B. acknowledges funding by the UK Science and Technology Facilities Council (STFC) grant no. ST/M001229/1.
V.F-G acknowledges funding from the National Aeronautics and Space Administration through the Exoplanet Research Program under Grant No. 80NSSC21K0394 (PI: S. Ertel)
This project is supported in part by the European Research Council (ERC) under the European Union's Horizon 2020 research and innovation programme (COBREX; grant agreement n° 885593).
F.Me. has received funding from the European Research Council (ERC) under the European Union's Horizon Europe research and innovation program (grant agreement No. 101053020, project Dust2Planets).
C.P. acknowledges funding from the Australian Research Council via FT170100040, DP180104235, and DP220103767.
\\

This research has made use of the NASA Exoplanet Archive, which is operated by the California Institute of Technology, under contract with the National Aeronautics and Space Administration under the Exoplanet Exploration Program.
This work made use of Astropy:\footnote{http://www.astropy.org} a community-developed core Python package and an ecosystem of tools and resources for astronomy \citep{astropy:2013, astropy:2018, astropy:2022}. 

\end{acknowledgements}

 \bibliographystyle{aa}
  \bibliography{elisabeth}

\begin{appendix} 
\section{Additional Information for Candidate Companions}
\label{app:cc_astrometry}

Five candidate companions were detected in  \citet{matthews2018} and re-detected in this work; these were assumed to be background objects in \citet{matthews2018}. We here confirm with common proper motion testing that these are indeed background objects. One new candidate companion is identified in this work towards the edge of the detector. This candidate is highly likely a background object, though additional data would be required to confirm this.

These candidates are all in the background-limited regime of the images, and a simple analysis of the IRDIS is sufficient to re-detect the companions and extract astrometry. We therefore derotated and co-added the cleaned image cubes from each epoch without applying PCA, and additionally co-added images from the two halves of the detector. We then determined the position of each candidate in each epoch by fitting the stellar PSF image to each companion, in a small box centered on the expected location of the candidate. The best-fitting position was converted into an on-sky astrometric acceleration, following the SPHERE User Manual\footnote{\href{https://www.eso.org/sci/facilities/paranal/instruments/sphere/doc.html}{https://www.eso.org/sci/facilities/paranal/instruments/sphere/doc.html}}. We propagated uncertainties from the detector properties and from the stellar PSF fit to calculate astrometric errors. Astrometry for each candideate at each epoch is listed in Table \ref{tab:candidate_astrometry}.

\begin{figure*}[t]
    \centering
    \includegraphics[width=\textwidth]{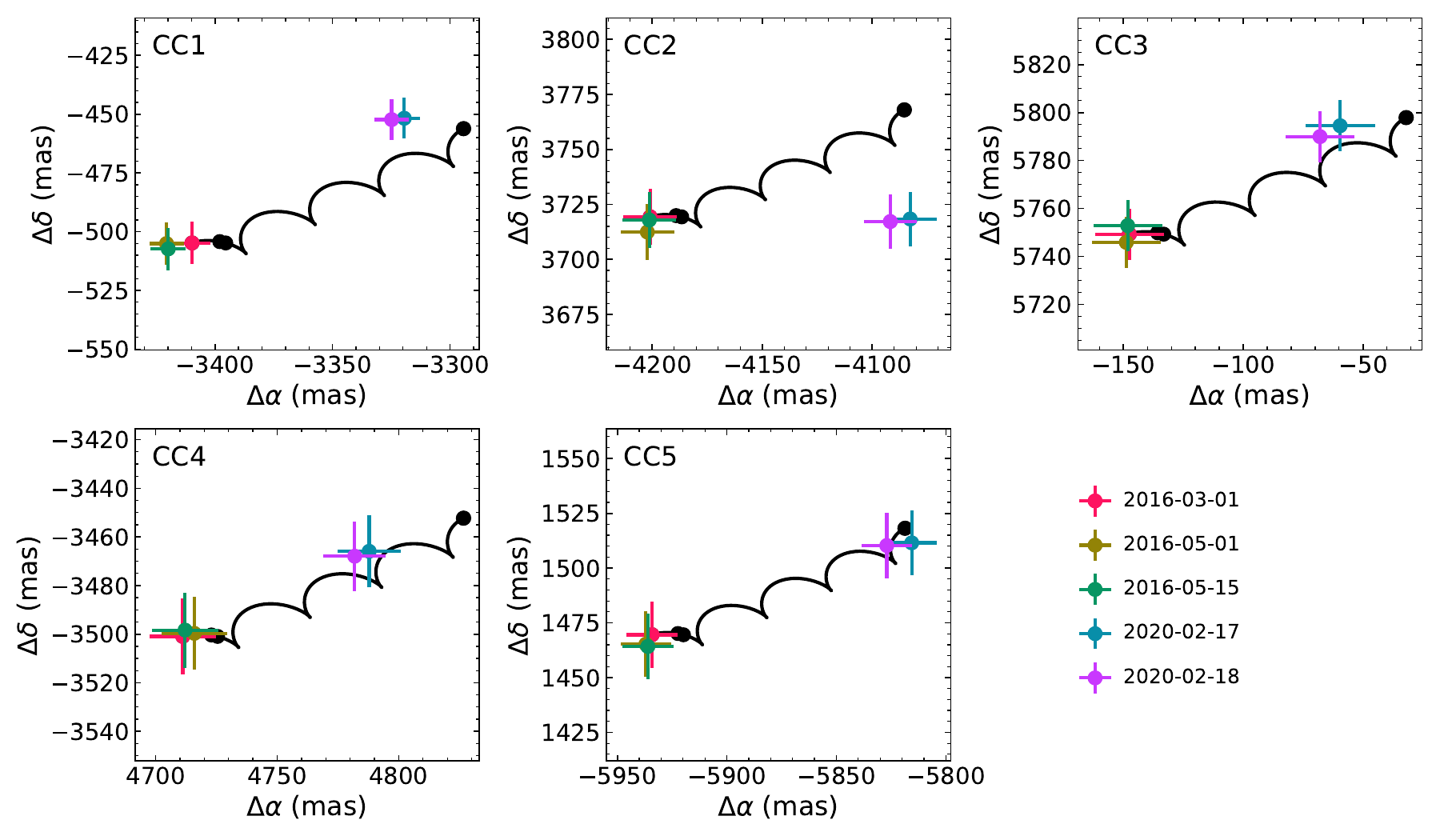}
    \caption{Companion astrometry overlaid on the background hypothesis for each of the candidate companions identified in 2016 data. In each case, coloured points represent the measured position and 1$\sigma$ uncertainty of the companion (also listed in Table \ref{tab:candidate_astrometry}); the black trace indicates the long-term motion of an infinitely distant background companion, with black circles indicating positions at the observation epoch. All companions move broadly in the direction expected for a background object, though there is some scatter -- implying a small proper motion of the background objects themselves. Considering the astrometry, as well as the wide separation (all $>$3.5\arcsec), all of these candidate companions are background objects. Companions CC2 and CC5 are also detected in the \gaia~catalog, further confirming their identity as background objects.}
    \label{fig:cpmtest}
\end{figure*}

For the five companions detected in both 2016 and 2020, we compared the measured astrometry to predicted movement of a stationary background object between 2016-2020. We confirmed that all five sources are indeed background stars, as shown in Fig.~\ref{fig:cpmtest}.  Two of these five sources have since been detected in \gaia. One of these has a parallax confirming it as a background star while the other has a G-H2 colour of $\sim$1.3, implying a star mid-G spectral type -- this is consistent with astrometric determination that this object is significantly more distant than HD\,112810.

\renewcommand{\arraystretch}{1.3}
\begin{table}[t]
    \centering
    \caption{Astrometry for all candidate companions to HD\,112810 identified in the VLT/SPHERE data.}
    \footnotesize
    \begin{tabular}{c|ccccc}
N & date & sep [mas] & $\sigma_\textrm{sep}$ [mas] & PA [$\deg$] & $\sigma_\textrm{PA}$ [$\deg$] \\
\hline
\hline
1 & 2016-03-01 & 3446.9 &  7.4 & 261.58 & 0.14 \\
1 & 2016-05-01 & 3457.7 &  6.8 & 261.60 & 0.14 \\
1 & 2016-05-15 & 3457.3 &  7.1 & 261.56 & 0.14 \\
1 & 2020-02-17 & 3350.1 &  6.0 & 262.25 & 0.14 \\
1 & 2020-02-18 & 3355.5 &  6.8 & 262.25 & 0.14 \\
\hline
2 & 2016-03-01 & 5610.8 &  9.7 & 311.52 & 0.14 \\
2 & 2016-05-01 & 5607.2 &  9.7 & 311.46 & 0.14 \\
2 & 2016-05-15 & 5610.1 &  9.7 & 311.51 & 0.14 \\
2 & 2020-02-17 & 5522.2 &  9.5 & 312.33 & 0.14 \\
2 & 2020-02-18 & 5528.1 &  9.5 & 312.25 & 0.14 \\
\hline
3 & 2016-03-01 & 5751.2 &  9.9 & 358.53 & 0.14 \\
3 & 2016-05-01 & 5747.8 &  9.9 & 358.52 & 0.14 \\
3 & 2016-05-15 & 5754.8 &  9.9 & 358.52 & 0.14 \\
3 & 2020-02-17 & 5794.8 & 10.0 & 359.41 & 0.14 \\
3 & 2020-02-18 & 5790.4 & 10.0 & 359.33 & 0.14 \\
\hline
4 & 2016-03-01 & 5869.5 & 10.3 & 126.62 & 0.14 \\
4 & 2016-05-01 & 5872.6 & 10.2 & 126.58 & 0.14 \\
4 & 2016-05-15 & 5868.7 & 10.3 & 126.59 & 0.14 \\
4 & 2020-02-17 & 5910.7 & 10.2 & 125.90 & 0.14 \\
4 & 2020-02-18 & 5907.0 & 10.2 & 125.95 & 0.14 \\
\hline
5 & 2016-03-01 & 6113.4 & 10.6 & 283.91 & 0.14 \\
5 & 2016-05-01 & 6115.1 & 10.6 & 283.86 & 0.14 \\
5 & 2016-05-15 & 6114.0 & 10.6 & 283.86 & 0.14 \\
5 & 2020-02-17 & 6008.8 & 10.4 & 284.57 & 0.14 \\
5 & 2020-02-18 & 6019.5 & 10.4 & 284.53 & 0.14 \\
\hline
6 & 2020-02-17 & 5217.2 &  9.5 &  98.78 & 0.14 \\
6 & 2020-02-18 & 5219.3 &  9.4 &  98.78 & 0.14 \\
\end{tabular}
 \tablefoot{For companions detected in both 2016 and 2020, we also plot the candidate astrometry, and the predicted astrometry for a background object, in Fig.~\ref{fig:cpmtest}; all companions are consistent with being background stars.}
    \label{tab:candidate_astrometry}
\end{table}
\renewcommand{\arraystretch}{0.769}

\section{Additional Figures for the Disk Modelling}
\label{app:extramodelling}

\begin{figure*}[t]
    \centering
    \includegraphics[width=1.\textwidth]{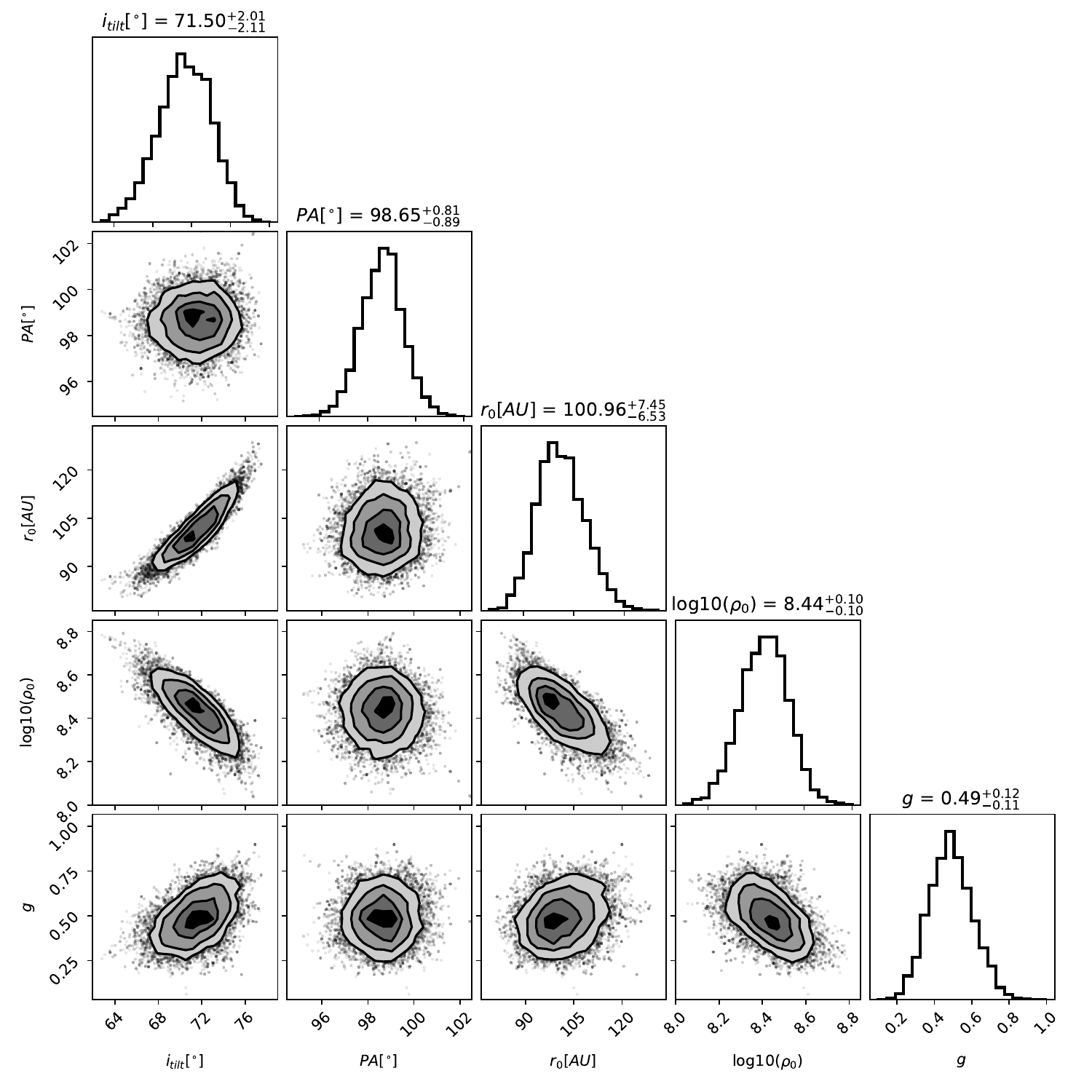}
    \caption{MCMC corner plot for the basic 5-parameter fit (Sect.~\ref{sec:5parfit}), highlighting the key correlations present in the data. Note that we treat $\rho_{0}$ as an arbitrary scaling factor. A clear correlation is observed between the disk inclination and characteristic radius, as is expected geometrically. The disk characteristic radius is also negatively correlated with the overall scaling factor $\rho_{0}$, which is a natural consequence of the $1/d^2$ scaling of scattered flux (see Equation \ref{eq:graterflux}).}
    \label{fig:model_corner_5par}
\end{figure*}

Here we include additional information for the disk modelling process, as described in Sect.~\ref{sec:model}. Specifically, Fig.~\ref{fig:model_corner_5par} shows the full corner plot for all the free parameters in the 5-parameter fit (Sect.~\ref{sec:5parfit}, parameters are $PA$, $i_\mathrm{tilt}$, r$_{0}$, $g$ and $\rho_{0}$). Corner plots are generated with \texttt{corner.py} \citep{corner}.

There is a clear correlation between the disk radius and the overall scaling of the disk: this is a natural result of the disk model setup (equation~\ref{eq:gratermain} above), where disk brightness falls with the square of distance from the star. To match the data, a larger disk needs a higher flux. These parameters are also correlated with the disk inclination, which is a natural result of the model disk geometry.

\begin{figure*}[t]
    \centering
    \includegraphics[width=1.\textwidth]{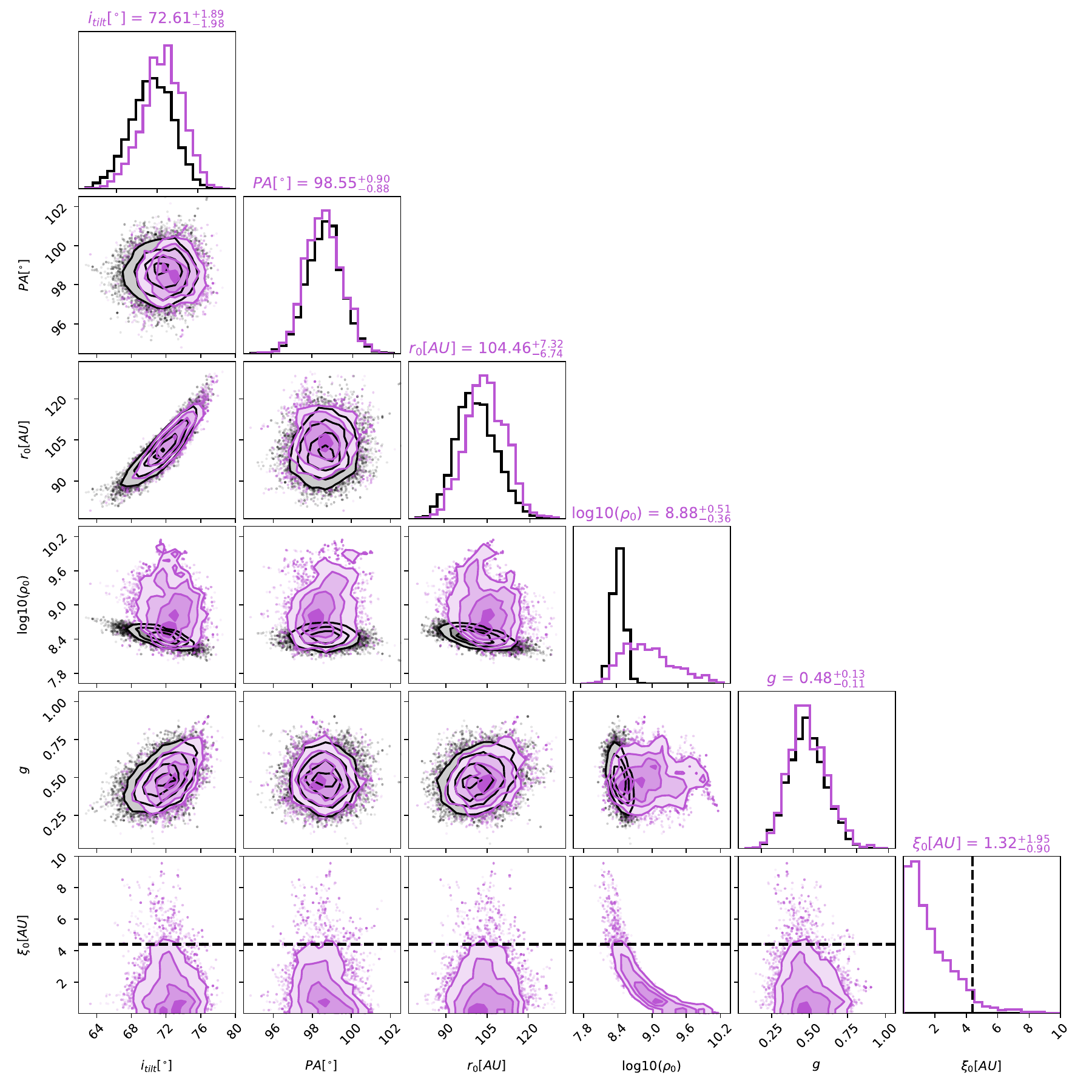}
    \caption{MCMC corner plot for the vertical structure fit (purple; Sect.~\ref{sec:diskvertical}) overlaid over the basic 5-parameter fit (black; Fig.~\ref{fig:model_corner_5par}). The fixed value of $\xi_0$ is indicated with a dashed black line. The $\xi_0$ posterior follows the prior, and the parameter is strongly correlated with $\rho_{0}$. All other parameters take similar values to the 5-parameter fit. We infer that the disk scale height $\xi_0$ cannot be usefully inferred from the in-hand data.}
    \label{fig:model_corner_6par}
\end{figure*}

\begin{figure*}[t]
    \centering
    \includegraphics[width=1.\textwidth]{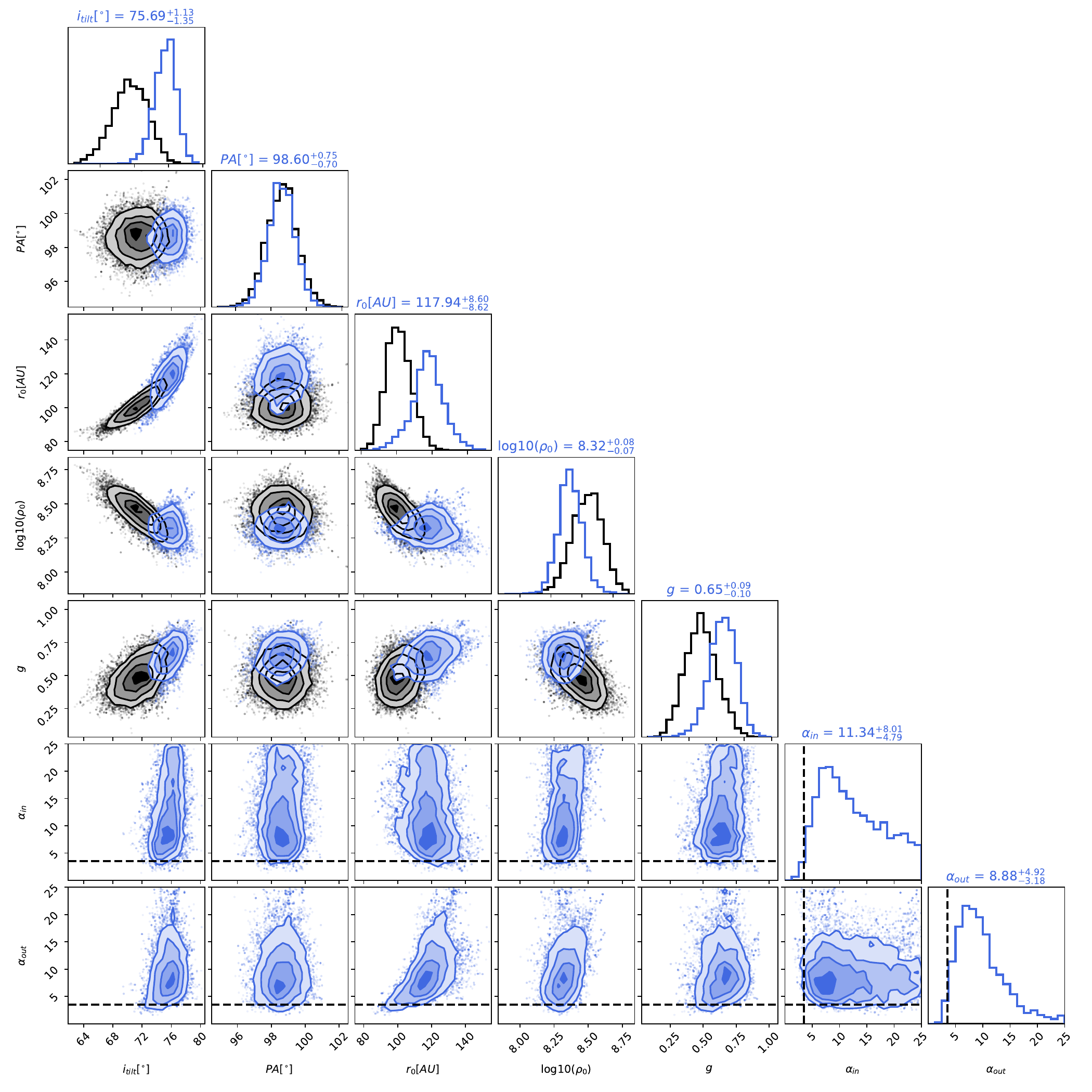}
    \caption{As Fig.~\ref{fig:model_corner_6par}, for the radial structure fit (blue; Sect.~\ref{sec:diskradial}). The dashed black lines indicate the fixed values of $\alpha_\mathrm{in}$ and $\alpha_\mathrm{out}$ for the 5-parameter fit. The posterior for $\alpha_\mathrm{out}$ is moderatley constrained; for $\alpha_\mathrm{in}$ values $\lesssim$10 are strongly disfavored, but higher values follow the prior. We infer that the disk profile is steep, but the extent of the sharpness is hard to constrain. Notably, the $r_0$ value is somewhat higher for this fit, as discussed in detail in Sect.~\ref{sec:diskradial}. The disk inclination value is also higher in this case, following the clear correlation of $r_0$ and $i_\mathrm{tilt}$ also highlighted in Fig.~\ref{fig:model_corner_5par}.}
    \label{fig:model_corner_7par}
\end{figure*}

Figs.~\ref{fig:model_corner_6par} and \ref{fig:model_corner_7par} show the full corner plot for the fits described in Sects.~\ref{sec:diskvertical} and \ref{sec:diskradial} respectively, overlaid on the 5-parameter fit. In each case we use dashed lines to indicate the values of fixed parameters in the 5-parameter fit. The $\xi_0$ posterior follows the prior, while the $\alpha_\mathrm{in}$ and $\alpha_\mathrm{out}$ posteriors strongly favor very steep power laws, corresponding to a narrow disk. $\alpha_\mathrm{in}$ and $\alpha_\mathrm{out}$ are correlated with $r_0$ and consequently also $i_\mathrm{tilt}$, and the best fit values for these two parameters are $\sim$1.5$\sigma$ higher than for the 5-parameter fit.

\end{appendix}

\end{document}